\documentclass[
reprint,
 amsmath,amssymb,
 aps,
pre,
]{revtex4-2}

\usepackage{graphicx}
\usepackage{dcolumn}
\usepackage{bm}
\usepackage{color}
\usepackage{comment} 

\newtheorem{dfn}{Definition}
\newtheorem{prop}{Proposition}

\newtheorem{remark}{Remark} 
\newtheorem{asm}{Assumption}

\newcommand{\sgn}{\mathrm{sgn}}
\newcommand{\rmi}{\mathrm{i}}

\begin{document}

\preprint{APS/123-QED}

\title{Sequential Retrieval in Dense Associative Memory: Asymptotic Dynamics and Storage Capacity}


\author{Mao Mishima}
 \email{9080506601mishima@g.ecc.u-tokyo.ac.jp}
\affiliation{Department of Physics, Faculty of Science, The University of Tokyo}
\affiliation{RIKEN Center for Advanced Intelligence Project }
\author{Ayaka Sakata}%
 \email{ayakasakata@is.ocha.ac.jp}
\affiliation{%
 Department of Information Science, 
 Ochanomizu University}
\affiliation{RIKEN Center for Advanced Intelligence Project }
\author{Kazushi Mimura}%
\email{mimura@hiroshima-cu.ac.jp}
\affiliation{Faculty of Information Sciences, Hiroshima City University}
\affiliation{RIKEN Center for Advanced Intelligence Project }
\affiliation{Department of Physics, Graduate School of Science, The University of Tokyo}
\date{\today}

\begin{abstract}

We study the retrieval dynamics of an $n$-body asymmetric dense associative memory model, in which sequential retrieval is implemented by shifting one pattern index in the higher-order Hebbian interaction. 
The model exhibits sequential retrieval of the stored patterns, and the retrieval state corresponds to a stable limit-cycle attractor. The storage capacity is defined as the maximum loading rate for which this limit cycle remains stable. Using generating functional analysis, we derive the dynamical equations describing the sequential retrieval state.
The generating functional analysis reveals that the retarded self-interaction term vanishes in the asymmetric model, and the resulting overlap dynamics is reproduced by an Amari--Maginu-type signal-to-noise analysis.
Also, we find that for the interaction order $n\ge 3$, the normalized storage capacity decreases monotonically with increasing interaction order, in contrast to symmetric Modern Hopfield models.

\end{abstract}
\maketitle


\section{\label{sec:level1}Introduction}

Associative memory models have provided a physically interpretable framework for studying information storage and retrieval in neural networks. 
In the seminal Hopfield model \cite{hopfield1982neural}, binary patterns are embedded as 
attractors of a recurrent dynamical system, and memory retrieval is described as relaxation toward one of these attractors. 
Subsequent studies established a close connection between associative memory models and physics for spin-glass \cite{amit1985spin}, 
enabling the analysis of storage capacity, retrieval phases, and memory stability using methods from statistical mechanics.

Recently, renewed interest in associative memory models has been stimulated by the development of dense associative memory or Modern Hopfield models with dramatically enhanced storage capacity \cite{krotov2016dense}.

One key ingredient underlying this capacity enhancement
is the introduction of higher-order interactions among neurons. 
In the conventional Hopfield model, the synaptic coupling is constructed from pairwise 
products of pattern components based on Hebbian rule, resulting in pairwise interactions between neurons.
In contrast, the Modern Hopfield model generalizes this to $n$-body interactions, constructed from products of $n$ patterns.
Such higher-order interactions are known to enhance the storage capacity, which scales as $O(N^{n-1})$ with the number of neurons $N$, while the prefactor depends nontrivially on the interaction order $n$ \cite{mimura2025dynamical}.

Beyond their roles as mathematical models of associative memory, modern Hopfield models have revealed a connection to modern machine learning architectures. 
The update rule of the continuous Modern Hopfield model \cite{ramsauer2021hopfield} is mathematically equivalent to the attention mechanism in the Transformer \cite{vaswani2017attention}, establishing a direct connection between associative memory retrieval and the core computation of large language models. 
This correspondence suggests that Hopfield-type models serve as a unifying framework bridging statistical physics, neural computation, and modern artificial intelligence.

Many modifications of the conventional Hopfield model have been proposed, including models with asymmetric synaptic couplings, motivated by the biological observation that synaptic connections in the brain are not necessarily symmetric \cite{derrida1987exactly}.
A particularly interesting example is the sequential associative memory, in which the asymmetric synaptic coupling is constructed by introducing a shift in the pattern index to the standard Hebbian rule. 
Under parallel updating of the neurons, the network dynamically recalls patterns as a temporal sequence rather than converging to static attractors \cite{sompolinsky1986temporal,coolen1992competition}. 
In general, the asymmetry of the synaptic coupling violates detailed balance, making conventional equilibrium statistical mechanical methods inapplicable.
However, it has been shown that the cyclic retrieval can be characterized as a limit cycle attractor, and its stability has been analyzed using generating functional analysis(GFA)  \cite{during1998phase}, yielding a larger storage capacity than that of the standard symmetric Hopfield model.

Higher-order interactions combined with pattern-shifting couplings were already studied before the recent resurgence of Modern Hopfield models
\cite{hamakawa2004some,miyajima2009higher}; we refer to this class of models as {\it modern sequence processing network}.
As in the pairwise case, these models exhibit sequential retrieval in the form of limit-cycle attractors, as observed numerically.

It was shown that, although the storage capacity scales as $O(N^{n-1})$, its coefficient decreases as the interaction order $n$ increases, in contrast to the symmetric case. 
More recent work studied the modern sequence processing network and derived its capacity for perfect recall using a signal-to-noise analysis \cite{chaudhry2023longsequencehopfieldmemory}.
However, the limit-cycle attractor and its phase structure have not yet been systematically characterized. It therefore remains unclear whether the predicted reduction in the coefficient of storage capacity is indeed an intrinsic property of the limit-cycle attractor and, if so, what mechanism underlies this reduction.

Recent work has used generating functional analysis to study the dynamics of symmetric dense associative memory \cite{mimura2025dynamical}. The resulting asymptotically exact single-unit description includes a retarded self-interaction that accounts for feedback from past states. Neglecting this term leads to an approximate scalar recursion for the macroscopic overlap.

In this paper, we address the questions raised above by considering a modern sequence-processing network under parallel updating dynamics and analyzing it using generating functional analysis~\cite{de1978dynamics}.
The model considered here can be viewed as an asymmetric sequence-processing counterpart of the symmetric dense associative memory described above, obtained by shifting one pattern index in the higher-order interaction.

We show that this modification qualitatively changes the effective dynamics, i.e., the shift removes the retarded self-interaction. For $n\geq 3$, the equal-time crosstalk variance is fixed only by the loading rate and interaction order, yielding an exact closed recursion for the one-time overlap despite possible temporal correlations in the effective noise.  

The resulting overlap recursion has the same form as that obtained in the symmetric model only after neglecting the retarded self-interaction, whereas here it follows directly from the generating functional analysis without that approximation.
Using this closed dynamics, we characterize the
sequential-retrieval limit cycle, its basin of attraction, and the
dependence of the critical loading rate on the interaction order. 

We also derive the large-$n$ asymptotics of the retrieval threshold
and discuss a formal connection to the single-site absolute-capacity
scale \cite{krotov2016dense, sakurai2026biasinducedcrossoverabsolutecapacity}.

This paper is organized as follows. 
Section \ref{sec:model} introduces the modern sequence processing network analyzed in this paper. Section \ref{sec:analysis} develops the GFA and reduces the microscopic dynamics to an effective single-unit process. Section \ref{sec:result} presents our main results: we derive the macroscopic retrieval dynamics in terms of a single macroscopic parameter and the stationary-state equations by considering the long-time limit of the dynamics on the limit cycle, and discuss their correspondence with the Amari--Maginu theory. Section \ref{sec:validation} compares these predictions with direct numerical simulations. Finally, Section \ref{sec:conclusion} concludes the paper with a summary and discussion.

\section{Model}
\label{sec:model}

The conventional modern Hopfield model is formulated through an energy function, whose energy decreases monotonically under asynchronous updating, so that the dynamics converges to a fixed-point attractor.
Since the model studied in this paper employs asymmetric couplings, such an energy function generally does not exist. We therefore begin by briefly reviewing the update rule of the conventional modern Hopfield model, which serves as the basis of the asymmetric extension.

\subsection{Update Rule of the Modern Hopfield Model}

The modern Hopfield model is a recurrent neural network that stores a large number of patterns as fixed-point attractors. Let $\bm{S}\in\{-1,+1\}^N$ denote the state of $N$ binary neurons. The energy function is defined as
\begin{align}
    H(\bm{S})
    =
    -\sum_{\mu=1}^{M}
    F\!\left(
        \sum_{i=1}^{N}\xi_i^\mu S_i
    \right),
\end{align}
where $\bm{\xi}^{\mu}\in\{-1,+1\}^{N}$ $(\mu=1,\ldots,M)$ are the stored patterns. 
Each component $\xi_i^\mu$ is independently drawn from $\{-1,+1\}$ with equal probability.

The key feature of the modern Hopfield model is the introduction of the nonlinear function $F$. 
Throughout this paper, we consider the widely used polynomial form
\begin{align}
    F(x)=\frac{x^n}{2N^{n-1}},
\end{align}
where $n(\geq 2)$ denotes the interaction order and the normalization factor $1/(2N^{n-1})$ is introduced only for convenience and does not affect the retrieval properties.
By introducing the nonlinear function $F$, the contribution of strongly correlated memory patterns is selectively amplified, while weaker overlaps are suppressed. 
This reduces interference among stored patterns and substantially improves the storage capacity.

The neuron states are updated according to the energy difference before and after flipping a single neuron. Denoting the state of neuron $i$ at time step $t$ by $S_i^{(t)}$, and retaining only the leading-order contribution, we obtain
\begin{align}
S_i^{(t+1)}
=
\sgn\!\left[
\sum_{\mu=1}^{M}
\xi_i^\mu
n
\left(
\frac{1}{N}
\sum_{j\ne i}
\xi_j^\mu
S_j^{(t)}
\right)^{n-1}
\right],
\end{align}
where $\sgn(x)$ denotes the sign function that takes 1 if $x\geq 0$ and $-1$ otherwise. 

For the subsequent dynamical analysis, we introduce an infinitesimal time-dependent external field $\theta_i^{(t)}$ to probe the response of the system. 
The update rule is then generalized as
\begin{align}
S_i^{(t+1)}
=
\sgn\!\left[
h_i^{(t)}\right],
\label{eq:def_neuron_update}
\end{align}
where $h_i^{(t)}$ is the local field acting on the $i$-th neuron at time step $t$, given by
\begin{align}
    h_i^{(t)} = n\sum_{\mu=1}^M 
	\xi_i^{\mu}  \biggl( \frac1N \sum_{j \ne i}^N \xi_j^\mu S_j^{(t)} \biggr)^{n-1} + \theta_i^{(t)}.
    \label{eq:localfield}
\end{align}
Expanding the \((n-1)\)-th power, we obtain another expression of the local field as 
\begin{align}
    h_i^{(t)}=n\sum_{j_1\neq i}^N\cdots\sum_{j_{n-1}\neq i}^NJ_{ij_1j_2\ldots j_{n-1}}S_{j_1}^{(t)}S_{j_2}^{(t)}\ldots S_{j_{n-1}}^{(t)}+\theta_i^{(t)},
    \label{eq:localfield_with_J}
\end{align}
with
\begin{align}
        J_{ij_1\ldots j_{n-1}}=\frac{1}{N^{n-1}}\sum_{\mu=1}^M\xi_{i}^{\mu}\xi_{j_1}^{\mu}\xi_{j_2}^{\mu}\ldots\xi_{j_{n-1}}^{\mu}.
    \label{eq:Jijk_symmetric}
\end{align}
For $n=2$, the local field given by Eq.~(\ref{eq:localfield_with_J}) reduces to that of the conventional Hopfield model with pairwise Hebbian couplings.

\subsection{Extension to Asymmetric Couplings: \\
Modern Sequence Processing Network}

We generalize the high-order interaction tensor of the modern Hopfield model \eqref{eq:Jijk_symmetric} by shifting one of the pattern indices by one, while the remaining $n-1$ indices retain the original pattern index:
\begin{align}
    J_{ij_1\ldots j_{n-1}}=\frac{1}{N^{n-1}}\sum_{\mu=1}^M\xi_{i}^{\mu+1}\xi_{j_1}^{\mu}\xi_{j_2}^{\mu}\ldots\xi_{j_{n-1}}^{\mu},
    \label{eq:Jijk}
\end{align}
where the pattern index $\mu$ is understood modulo $M$ \cite{during1998phase,hamakawa2004some,miyajima2009higher}.
This asymmetric interaction induces sequential retrieval, $\bm{\xi}^\mu\to\bm{\xi}^{\mu+1}\to\bm{\xi}^{\mu+2}\to\cdots$, when the neurons are updated in parallel.
The mechanism of sequential retrieval can be seen from the local field at time $t$ under the interaction \eqref{eq:Jijk}, which is given by
\begin{align}
    h_i^{(t)} = \sum_{\mu=1}^M 
	\xi_i^{\mu+1} n \biggl( \frac1N \sum_{j \ne i}^N \xi_j^\mu S_j^{(t)} \biggr)^{n-1} + \theta_i^{(t)}.
    \label{eq:local_field_asymmetric}
\end{align}
When the overlap between the state $\bm{S}^{(t)}$ and the pattern $\bm{\xi}^{\mu}$ becomes large, 
the contribution of the subsequent pattern $\bm{\xi}^{\mu+1}$ to the local field at time $t$ is enhanced, which drives the next state $\bm{S}^{(t+1)}$ toward $\bm{\xi}^{\mu+1}$.

The storage capacity is characterized by the stability of the limit-cycle attractor associated with sequential retrieval \cite{during1998phase,hamakawa2004some,miyajima2009higher}. 
In the following sections, we investigate how the stability of this attractor and the resulting storage capacity depend on the interaction order $n$.

\section{Analysis}
\label{sec:analysis}

The dynamics defined by the transition rule \eqref{eq:def_neuron_update} does not, in general, admit an energy function under the asymmetric interaction tensor $J$. Consequently, conventional equilibrium statistical mechanics based on an energy function is not directly applicable. We therefore follow the generating-functional analysis developed for symmetric dense associative memory in Ref.~\cite{mimura2025dynamical}, modifying it to account for the shifted pattern index in the interaction tensor.

\subsection{Generating Functional Analysis}

In the GFA formalism, we consider the path probability of the microscopic states over a finite time interval, from the initial condition to a final time.
The generating functional is introduced to evaluate dynamical observables, such as the overlap between the microscopic state vector and the retrieved memory pattern, from the path probability.

In case of parallel updating, the transition probability from $\bm{S}^{(t)}$ to $\bm{S}^{(t+1)}$ is given by
\begin{align}
	p[\bm{S}^{(t+1)}|\bm{S}^{(t)}] = \prod_{i=1}^N \delta [ S_i^{(t+1)}; \sgn(h_i^{(t)}) ], 
\end{align} 
where $\delta [;]$ denotes the Kronecker delta and $h_i^{(t)}$ depends on $\bm{S}^{(t)}$.
We define path probability, which describes the joint probability distribution of microscopic states from time step $0$ to $T$.
Introducing the shorthand $\bm{S}^{0:T}=(\bm{S}^{(0)},\ldots,\bm{S}^{(T)})$, it is given by

\begin{align}
	p[\bm{S}^{0:T}]
    = 
	p[\bm{S}^{(0)}] \prod_{t=0}^{T-1} p[\bm{S}^{(t+1)}|\bm{S}^{(t)}].
	\label{eq:def_path_probability}
\end{align}

In the present model, the primary quantity of interest is the overlap between the microscopic state and a stored pattern,
\begin{align}
    m^\nu{(t)}
    &=\frac1N\sum_{i=1}^{N}\xi_i^\nu S_i^{(t)},
\end{align}
together with its average 
\begin{align}
    \overline{m^\nu}{(t)}
    &=
    \mathbb{E}_{\{\bm{\xi}^{\mu}\}}
    \left[
    \left\langle m^\nu{(t)}\right\rangle
    \right],
    \label{eq:def_m_average}
\end{align}
where $\langle\cdots\rangle$ denotes the dynamical average according to the path probability \eqref{eq:def_path_probability} under fixed patterns and $\mathbb{E}_{\{\bm{\xi}^{\mu}\}}$ denotes the disorder average over the $M$-stored patterns $\{\bm{\xi}^1,\cdots\bm{\xi}^M\}$. 
In order to analyze the dynamics of the overlap within the GFA framework, we introduce the generating functional associated with the path probability:
\begin{dfn}[Generating Functional]
Let us denote generating variables as $\bm{\psi}^{(t)}\in\mathbb{R}^N$ for $t\in\{0,1,\ldots,T-1\}$. Generating functional is defined as 
\begin{align}
\bar{Z}[\bm{\psi}]
=
\mathbb{E}_{\{\bm{\xi}^{\mu}\}}
\biggl[
\sum_{\bm{S}^{0:T-1}}
p[\bm{S}^{0:T-1}]
\exp\biggl(
-i\sum_{t=0}^{T-1}
\bm{S}^{(t)}\cdot\bm{\psi}^{(t)}
\biggr)
\biggr],
\label{eq:def-Z}
\end{align}
where $\bm{\psi}=\{\bm{\psi}^{(t)}\}_{t=0}^{T-1}$.
\end{dfn}

The generating functional allows us to derive macroscopic dynamical order parameters, such as the overlap,
by differentiating with respect to $\bm{\psi}$ and then taking $\psi_i^{(t)}\to 0$ for all $i$ and $t$.
We therefore calculate the generating functional to characterize the retrieval dynamics.
To enforce the definition of the local field in \eqref{eq:local_field_asymmetric}, we introduce the identity 
\[
    1=\int dh_i\delta\left(h_i^{(t)}-\sum_{\mu=1}^M 
	\xi_i^{\mu+1} n( \frac1N \sum_{j \ne i}^N \xi_j^\mu S_j^{(t)})^{n-1} - \theta_i^{(t)}\right),
\]
for all $i$ and $t$, where $\delta(\cdot)$ denotes the Dirac delta function.
We then obtain
\begin{align}
	\nonumber
    \bar{Z}[\bm{\psi}]
    &=\mathbb{E}_{\{\bm{\xi}^{\mu}\}}
    \sum_{\bm{S}^{(0:{T-1})}}
	\int_{\mathbb{R}^{NT}} d\bm{h}
    e^{ - i \sum_{t=0}^{T-1} \bm{S}^{(t)} \cdot \bm{\psi}^{(t)} }\\
    \nonumber
	&\times p[\bm{S}^{(0)}]
	\biggl( 
		\prod_{t=0}^{T-1} \prod_{j=1}^N 
		\delta [S_j^{(t+1)} ; \mathrm{sgn}(h_j^{(t)}) ] 
	\biggr) \\
     &\times 
        \prod_{t=0}^{T-1} \! \prod_{j=1}^N \delta\biggl(h_j^{(t)}\!\!-\!\!\sum_{\mu=1}^M 
	\xi_j^{\mu+1} n( \frac1N \!\sum_{k \ne j}^N \xi_k^\mu S_k^{(t)})^{n-1} \!\!- \!\theta_j^{(t)}
    \biggr),
\end{align}
where $d\bm{h}$ is the set of $dh_i^{(t)}$ for all $i\in\{1,\ldots,N\}$ and $t\in\{0,\ldots,T-1\}$.

We exploit the characteristic
dynamics induced by the pattern-shifting interaction.
As discussed in previous studies \cite{hamakawa2004some,miyajima2009higher}, the interaction tensor
\eqref{eq:Jijk} generates sequential retrieval dynamics, in which the
stored patterns are recalled one after another.
As in the case of $n=2$, 
the stored patterns are retrieved sequentially, $\bm{\xi}^{t}\to \bm{\xi}^{t+1}\to\bm{\xi}^{t+2}\to\cdots$.
Motivated by this observation, we introduce the following assumption.

\begin{asm}[Sequential retrieval assumption]
Without loss of generality, we label the condensed pattern at time
$t$ by the same index $t$.
At each time step, one stored pattern is assumed to be
condensed, in the sense that its overlap with the system state is of
order $O(1)$, whereas the overlaps with all other stored patterns are
of order $O(N^{-1/2})$.
Accordingly, we define
\begin{align}
    m^{(t)}=m^{t}(t)=\frac{1}{N}\sum_{i=1}^N\xi_i^tS_i^{(t)},
    \label{eq:def_mt}
\end{align}
which is the overlap with the pattern condensed at time t, and serves as the order parameter characterizing the sequential retrieval dynamics.
\end{asm}
Here, we consider the thermodynamic limit $N\to\infty$, where $M\to\infty$ as well. 
Therefore, we can assume $T<M$ for sufficiently large $M$, so that there exist patterns with $\mu>T$ that have not yet been retrieved.

Following the above discussion, we separate the contributions in $\overline{Z}[\bm{\psi}]$ into those from the retrieved pattern and those from the non-retrieved patterns. Introducing the Fourier representation of the Dirac delta function for ${h_i^{(t)}}$ with conjugate variables ${\widehat{h}_i^{(t)}}$, the generating functional can be written as
\begin{align}
\nonumber
\bar{Z}[\bm{\psi}]
=\!\!&
\sum_{\bm{S}^{0:T}}
\!\int \!d\bm{h}d\hat{\bm{h}}
p[\bm{S}^{(0)}]
\prod_{t=0}^{T-1}
\prod_{i=1}^{N}
\delta\left[S_i^{(t+1)};\mathrm{sgn}\left(h_i^{(t)}\right)\right]\\
&\times\exp\left(-i\sum_{t=0}^{T-1}\bm{S}^{(t)}\cdot\bm{\psi}^{(t)}\right)
\nonumber\\
&\times\exp\biggl[i\sum_{t=0}^{T-1}\sum_{i=1}^{N}\hat h_i^{(t)}
\left(h_i^{(t)}-\theta_i^{(t)}\right)
\biggr]\Xi(\bm{S}^{0:T}), 
\label{eq:noise-in-Z}
\end{align}
where 

\begin{align}
\nonumber
    &\Xi(\bm{S}^{0:T})\\
    &\!=\!\mathbb{E}_{\{\bm{\xi}^{\mu}\}}\!
\biggl[\exp\biggl(\!-inN\!\!\sum_{t=0}^{T-1}\!k^{t+1,t}\!\!\left(m^{(t)}\right)^{\!n-1}
\hspace{-0.5cm}-i\!\sum_{\mu=1}^M\!\!\!\!X_\mu
\!+\!O(\frac{1}{N})
\biggr)
\!\biggr]
\label{eq:Xi}
\end{align}
and 
\begin{align}
    X_\mu&=n\sum_{\substack{t=0\\t\ne \mu}}^{T-1}\sum_{i=1}^{N}\hat h_i^{(t)}
\xi_i^{\mu+1}
\biggr(\frac1N\sum_{j\ne i}^{N}\xi_j^{\mu}S_j^{(t)}\biggr)^{n-1}\label{eq:def_X_mu}\\
k^{\mu,t}
    &=\frac1N\sum_{i=1}^{N}\xi_i^\mu \hat h_i^{(t)},\label{eq:def_k}
\end{align}
and $d\widehat{\bm{h}}$ is the set of $d\hat{h}_i^{(t)}$.

As shown in Appendix \ref{sec:app_prop_A}, 
$\overline{Z}[\bm{\psi}]$ 
can be described by the following quantities in addition to $\{m^{(t)}\}_{t=0}^T$:
\begin{align}
    q(t,t')
    &=\frac1N\sum_{i=1}^{N}S_i^{(t)} S_i^{(t')},\label{eq:def_q}\\
    Q(t,t')
    &=\frac1N\sum_{i=1}^{N}\hat h_i^{(t)} \hat h_i^{(t')},\label{eq:def_Q}\\
    K(t,t')
    &=
    \frac1N\sum_{i=1}^{N}S_i^{(t)} \hat h_i^{(t')}.
\end{align}
\subsection{Disorder average and saddle-point equations}
We introduce the matrix notation 
$\underline{{\cal A}}\in\mathbb{R}^{T\times T}$ 
for quantities with two time indices ${\cal A}^{(t,t')}$, 
where the $(a,b)$ component of $\underline{{\cal A}}$ is given by ${\cal A}^{(a-1,b-1)}$.
Introducing the conjugate variables
$\underline{\hat{\bm{m}}},\underline{\hat{\bm{k}}},\underline{\hat{q}},\underline{\hat{Q}},\underline{\hat{K}}$,
and denoting $\underline{\bm{a}}_i=[a_i^{(0)},\cdots,a_i^{(T-1)}]^\top$, we obtain the following proposition.

\begin{prop}
\label{prop:Z_form}
For $n\geq 3$, averaging the generating functional over all the stored patterns gives
    \begin{align}
	   \bar{Z}[\bm{\psi}]
		&= 
		\int
		d\bm{m} d\hat{\bm{m}}
		d\bm{k} d\hat{\bm{k}}
		d\bm{q} d\hat{\bm{q}}
		d\bm{Q} d\hat{\bm{Q}}
		d\bm{K} d\hat{\bm{K}}  \notag \\
		&\times\exp \biggl[ N(\Psi+\Phi+\Omega) + O(\log N) \biggr].
		\label{eq:bar-Z-in-lemma}
    \end{align}
	Here, 
    \begin{align}
		\Psi
		=& 
		i \sum_{t=0}^{T-1} \biggl\{ 
		\hat{m}^{(t)} m^{(t)} + 
		\hat{k}^{(t)} k^{(t)} - 
		k^{(t)} ~ n ( m^{(t)} )^{n-1}
		\biggr\} 		\notag \\
        +& i  \biggl\{ 
		\mathrm{Tr}\left(\underline{\hat{q}}^\top\underline{q}\right) + 
		\mathrm{Tr}\left(\underline{\hat{Q}}^\top\underline{Q}\right) + 
		\mathrm{Tr}\left(\underline{\hat{K}}^\top\underline{K} \right)\biggr\} \\
		\Phi
		=& 
		\frac1N\sum_{i=1}^N
		\log \!{\cal Z}_i(\widehat{\cal Q},\underline{\bm{\psi}}_i)\\
		\Omega
		=& 
		-\frac{n^2M}{2N^{n-1}} 
        \mathrm{Tr}(\underline{W}^\top\underline{Q})
		+O(N^{-1}). 
        \label{eq:Omega}
    \end{align}
The operator $\mathrm{Tr}(\cdot)$ denotes the matrix trace, and
$\underline{W}\in\mathbb{R}^{T\times T}$ is the matrix representation of
\begin{align}
\label{eq:def_W}
W(t,t^\prime)=\sum_{k=0}^{n-1} A(n-1,k) 
		( q^{(t,t')} )^k,
\end{align}
where 
\begin{align}
    A(\ell,k)
&=
\binom{\ell}{k}^{2} k! \, B(\ell-k)^2 ,\label{eq:def_A}\\
B(m)
&=
\mathbf{1}_{m:\mathrm{even}}\,(m-1)!!\label{eq:def_B}
\end{align}
with $\mathbf{1}_{m:\mathrm{even}}$ denoting the indicator function that is equal to one when $m$ is even and zero otherwise, and $!!$
denoting the double factorial.

Furthermore, let ${\cal \widehat{Q}}=\{\widehat{\bm{m}},\widehat{\bm{k}},\widehat{\bm{q}},\widehat{\underline{Q}},\widehat{\underline{K}}\}$, then
\begin{align}
    &{\cal Z}_i(\widehat{\cal Q},\underline{\bm{\psi}}_i)=\int d\underline{\bm{h}}_id\widehat{\underline{\bm{h}}}_i\sum_{\underline{\bm{S}}_i}\mathbb{E}_{\bm{\xi}}\biggl[\omega_i(\underline{\bm{S}}_i,\underline{\bm{h}}_i,\widehat{\underline{\bm{h}}}_i;\widehat{\cal Q},\underline{\bm{\psi}}_i)\biggr]
    \label{eq:Z_ave_prop1}\\
\nonumber
&\omega_i(\underline{\bm{S}}_i,\underline{\bm{h}}_i,\widehat{\underline{\bm{h}}}_i;\widehat{\cal Q},\underline{\bm{\psi}}_i)=p[S_i^{(0)}]
			\prod_{t=0}^{T-1}  
			\delta [S_i^{(t+1)} ; \mathrm{sgn}(h_i^{(t)}) ]\\
    &\hspace{2.0cm}\times\exp \biggl(
		-i  {\cal H}_{i}(\underline{\bm{S}}_i,\underline{\bm{h}}_i,\widehat{\underline{\bm{h}}}_i;\widehat{\cal Q},\underline{\bm{\psi}}_i)\biggl)
        \label{eq:def_omega}\\
\nonumber
        &{\cal H}_{i}=
		\bm{S}_i^\top\underline{\hat{q}}\bm{S}_i + 
		\hat{\bm{h}}_i^\top\underline{\hat{Q}}\hat{\bm{h}}_i + 
		\hat{\bm{h}}_i^\top\underline{\hat{K}}^\top\bm{S}_i+\underline{\bm{\psi}}_i^\top\underline{\bm{S}}_i  \\
        &-\sum_{t=0}^{T-1}\hat{h}_i^{(t)}
		\{ h_i^{(t)} - \hat{k}^{(t)}\xi_i^{t+1} - \theta_i^{(t)} \} +  \sum_{t=0}^{T-1} S_i^{(t)} \hat{m}^{(t)} \xi_i^{t}.
        \label{eq:eff_H}
\end{align}

\end{prop}

\begin{remark}
The absence of a retarded self-interaction can already be traced to the structure of \eqref{eq:Omega}. In the symmetric dense associative memory studied in Ref.~\cite{mimura2025dynamical}, the disorder average generates an additional contribution proportional to
\(K^{(t,t')}K^{(t',t)}\), which gives rise to the retarded self-interaction in the effective single-unit dynamics; see Sec.~\ref{sec:no_retarded_interaction}.
In the present sequence-processing model, this contribution is absent because the shifted output pattern \(\xi_i^{\mu+1}\) is statistically independent of the input-pattern factors \(\{\xi_j^\mu\}\).  Consequently, the disorder average over the output-pattern variables eliminates the contractions responsible for the retarded self-interaction.

\end{remark}

The scaling of the number of memory patterns $M$ is determined by the balance between the signal originating from the retrieved pattern and the noise originating from the non-retrieved patterns. From \eqref{eq:Omega}, $M$ must scale as $O(N^{n-1})$ in order for the generating functional to be of order $\exp(O(N))$. Therefore, we set
\begin{align}
    M=\alpha_n N^{n-1},
\end{align}
and define 
\begin{align}
    R^{(t,t^\prime)}=n^2\alpha_nW{(t,t^\prime)}.
    \label{eq:time_covariance}
\end{align}

We can apply the saddle-point method to Eq.~\eqref{eq:bar-Z-in-lemma} for sufficiently large $N$, namely
\begin{align}
    \overline{Z}[\bm{\psi}]=\exp\biggl[N\mathop{\mathrm{extr}}_{{\cal Q},\widehat{\cal Q}}(\Psi+\Phi+\Omega)\biggr],
\end{align}
where $\mathrm{extr}_{{\cal Q},\widehat{\cal Q}}$ denotes the extremization with respect to ${\cal Q}:=\{\bm{m},\bm{k},\bm{q},\underline{Q},\underline{K}\}$ and
$\widehat{\cal Q}$.
At the saddle point, which we denote with $*$, the following relationships should hold:
\begin{align}
	m^{*(t)} &= \frac{1}{N}\sum_{i=1}^N\langle\xi_i^t S_i^{(t)} \rangle_i\label{eq:m_at_saddle}\\ 
    q^{*(t,t^\prime)} &=\frac{1}{N}\sum_{i=1}^N\langle S_i^{(t)}S_i^{(t^\prime)} \rangle_i\label{eq:q_at_saddle}\\ 
    k^{*(t)}&=\frac{1}{N}\sum_{i=1}^N\langle \xi_{i}^{t+1}\widehat{h}_i^{(t)}\rangle_i\label{eq:k_at_saddle}\\
    Q^{*(t,t^\prime)}&=\frac{1}{N}\sum_{i=1}^N\langle\widehat{h}_i^{(t)}\widehat{h}_i^{(t^\prime)}\rangle_i\label{eq:Q_at_saddle}\\
    K^{*(t,t^\prime)}&=\frac{1}{N}\sum_{i=1}^N\langle S_i^{(t)}\widehat{h}_i^{(t^\prime)}\rangle_i\label{eq:K_at_saddle}
\end{align}
where the average $\langle (\cdots) \rangle_i$ is referred to as a {\it single-unit measure}
\begin{align}
\nonumber
    \langle f(&\underline{\bm{S}}_i,\bm{\xi},\underline{\bm{h}}_i,\widehat{\underline{\bm{h}}}_i)\rangle_i
    =\lim_{\underline{\bm{\psi}}_i\to \bm{0}}\frac{1}{{\cal Z}_i(\widehat{\cal Q}^*,\underline{\bm{\psi}}_i)}\sum_{\underline{\bm{S}}_i}\int d\underline{\bm{h}}_id\underline{\widehat{\bm{h}}}_i \\
    &\times \mathbb{E}_{\bm{\xi}}\biggl[f(\underline{\bm{S}}_i,\bm{\xi},\underline{\bm{h}}_i,\widehat{\underline{\bm{h}}}_i)\omega_i(\underline{\bm{S}}_i,\underline{\bm{h}}_i,\widehat{\underline{\bm{h}}}_i;\widehat{\cal Q}^*,\underline{\bm{\psi}}_i)\biggr].
    \label{eq:single_unit_measure}
\end{align}

As shown in Appendix~\ref{sec:app_saddle}, the saddle-point value $m^{*(t)}$ corresponds to $\overline{m}^t(t)$ defined by Eq.~\eqref{eq:def_m_average}. Hereafter, we omit the superscript $*$ indicating the saddle-point value. Unless otherwise stated, all quantities below are evaluated at the saddle point.

\section{Main Results}
\label{sec:result}

\subsection{Absence of retarded self-interaction in effective dynamics}
\label{sec:no_retarded_interaction}

Following the calculation shown in Appendix \ref{sec:app_prop_B}, by performing the integrals over $\bm{h}$ and $\hat{\bm{h}}$ at the saddle of ${\cal Q}$ and $\widehat{\cal Q}$
and taking the expectation over the patterns, we obtain the following Proposition \ref{prop:effective_Gaussian}.
\begin{prop}
\label{prop:effective_Gaussian}
Let us consider a site $i$ for which
$\xi_i^{(t+1)}=\xi^{t+1}$ and $\theta_i^{(t)}=\theta^{(t)}$.
For any observable 
$f(\underline{\bm{S}},\underline{\bm{\xi}})$, its single-unit measure $\langle f(\underline{\bm{S}})\rangle_i$ for $n\geq 3$ is described by

    \begin{align}
    \nonumber
	   \langle f(\underline{\bm{S}}&,\underline{\bm{\xi}}) \rangle_i =
	   \mathbb{E}_{\underline{\bm{\xi}}}\biggl[ \int \mathcal{D}\underline{\bm{v}}\sum_{\underline{\bm{S}}} f(\underline{\bm{S}},\underline{\bm{\xi}})
	   p[S^{(0)}] \\
       &\times\prod_{t=0}^{T-1}
	   \delta [ 
	   S^{(t+1)} ; 
	   \sgn ( \xi^{t+1} n(m^{(t)})^{n-1} 
	    \!\!+\! v^{(t)} \!+\! \theta^{(t)} )
	   ]\biggr], 
    \label{eq:def_average}
    \end{align}
    where
    \begin{align}
        {\cal D}\underline{\bm{v}}=\frac{d\underline{\bm{v}}}{\sqrt{(2\pi)^T|R|}}\exp\left(-\frac{1}{2}\underline{\bm{v}}^\top R^{-1}\underline{\bm{v}}\right).
    \end{align}

\end{prop}

The random vector $\underline{\bm{v}}$ in eq.\eqref{eq:def_average}, whose components correspond to different time steps, represents the crosstalk noise generated by the non-condensed patterns. Its temporal correlations are described by the covariance matrix $\underline{R}$.

The expression \eqref{eq:def_average} indicates that the dynamics of a single spin variable is statistically equivalent to
\begin{align}
S^{(t+1)}\overset{d}{=} 
\operatorname{sgn}
\left(\xi^{t+1} n(m^{(t)})^{n-1}+v^{(t)}+\theta^{(t)}\right),
\label{eq:S_effective_dynamics}
\end{align}
and thus is governed by the one-time overlap $m^{(t)}$ and Gaussian noise.

It is instructive to compare the effective single-unit dynamics \eqref{eq:S_effective_dynamics} with that of the symmetric dense associative memory derived in Ref. \cite{mimura2025dynamical}. In the notation of the present paper, the latter takes the form 
\begin{align}
S^{(t+1)}
\overset{d}{=}
\sgn\biggl[
\xi^{t+1} n\bigl(m^{(t)}\bigr)^{n-1}
+ (\Gamma S)^{(t)}
+ v^{(t)}
+ \theta^{(t)}
\biggr],
\end{align}
where the matrix $\Gamma$ denotes the retarded self-interaction kernel. Its contribution can be written as $(\Gamma S)^{(t)} = \sum_{t'<t}\Gamma(t,t')S^{(t')}$, and represents the feedback of past single-unit states onto the current effective field through the recurrent network. Thus, a nonzero $\Gamma$ introduces an explicit dependence of the effective dynamics on the history of the unit. In contrast, such retarded self-interaction term does not appear in \eqref{eq:S_effective_dynamics} for the present asymmetric model. 
It should be noted, however, that the absence of the retarded self-interaction alone is not sufficient for the overlap dynamics to be described solely in terms of the one-time overlap. For the 2-body sequence-processing model, additional crosstalk-noise contributions remain at leading order, preventing the dynamics from being determined solely by the one-time overlap~\cite{during1998phase}. 
For $n\ge3$, the regime considered in this work, these higher-order contributions become negligible in the thermodynamic limit and do not affect the leading-order effective dynamics. See Appendix ~\ref{sec:app_prop_A} for details.

\subsection{Closure of the dynamics in a single order parameter}

At this stage, the remaining site dependence appears only through the components of the condensed sequence patterns and the site-dependent external fields. 
This dependence can be eliminated for $m^{(t)}$ and $q^{(t,t^\prime)}$, 
by considering the case $\theta_i^{(t)}=0$ for all $i$ and $t$, which we focus on hereafter.
From \eqref{eq:def_average}, we obtain
\begin{align}
    &\langle\xi_i^{t+1}S_i^{(t+1)}\rangle_i=\mathbb{E}_{\underline{\bm{\xi}}_i}
	   \biggl[ \int \mathcal{D}\underline{\bm{v}}\xi_i^{t+1}\sgn ( \xi_i^{t+1} n(m^{(t)})^{n-1} 
	    \!\!+\! v^{(t)} )
	   \biggr]\notag\\
       &=\mathbb{E}_{\xi_i^{t+1}}
	   \biggl[ \int \frac{dv}{\sqrt{2\pi R^{(t,t)}}}\exp\biggl(-\frac{{v^{(t)}}^2}{2R^{(t,t)}}\biggr)\notag\\
       &\hspace{2.0cm}\times\sgn ( n(m^{(t)})^{n-1} 
	    \!\!+\! \xi_i^{t+1}v^{(t)} )\biggr]\notag\\
       &=\int \frac{dv}{\sqrt{2\pi R^{(t,t)}}}\exp\left(-\frac{{v^{(t)}}^2}{2R^{(t,t)}}\right)\sgn \biggl( n(m^{(t)})^{n-1} 
	    \!\!+\! v^{(t)} \biggr),
\end{align}
where we have used the change of variables $v^{(t)}\to \xi_i^{t+1}v^{(t)}$, which leaves the Gaussian integral invariant.

Therefore, $\langle\xi_i^{t+1}S_i^{(t+1)}\rangle_i$ loses $i$-dependency, and 
at the saddle point, we obtain
\begin{align}
\nonumber
m^{(t+1)}&=\int \frac{dv^{(t)}}{\sqrt{2\pi R^{(t,t)}}}\exp\biggr(-\frac{{v^{(t)}}^2}{2R^{(t,t)}}\biggl)\\
&\hspace{0.5cm}\times\operatorname{sgn}
\left(n(m^{(t)})^{n-1}+v^{(t)}\right).
\label{eq:m_time_evolution}
\end{align}

Hereafter, we naturally set $R^{(t,t)}$ as $R_0$ for any $t$.
Since \(S_i^{(t)}=\pm1\), (\ref{eq:def_q}) gives \(q^{(t,t)}=1\) for all \(t\).
Hence, using (\ref{eq:def_W}), (\ref{eq:time_covariance}) and (\ref{eq:app_Z_derivative_psi2}) shown in Appendix \ref{sec:app_prop_B}, we obtain 
\begin{align}
R_0
=n^2\alpha_n\sum_{k=0}^{n-1}A(n-1,k)=
n^2\alpha_n{(2n-3)!!}
\end{align}
As a consequence of the discussion toward the end of the previous subsection, the one-time overlap dynamics closes into a scalar recursion depending only on the current overlap, without dependence on any other physical quantities.
Remarkably, this recursion has the same form as that obtained in Ref. \cite{mimura2025dynamical} by neglecting the retarded self-interaction, i.e., by setting $\Gamma=O$, which was an approximation in the symmetric model. Here, the corresponding scalar recursion follows directly from the exact generating functional analysis.

\subsection{Stationary state}

 At the stationary state, we can set \(m^{(t+1)}=m^{(t)}=m\) in (\ref{eq:m_time_evolution}), therefore the order parameter satisfies the self-consistent equation
\begin{align}
m&=\int Dz~\operatorname{sgn}
\left(nm^{n-1}+z\sqrt R_0\right),
\label{eq:m_final}
\end{align}
and

\begin{align}
m
=
\operatorname{erf}
\left(
\frac{n m^{n-1}}
{\sqrt{2R_0}}
\right)
=
\operatorname{erf}
\left(
\frac{m^{n-1}}
{\sqrt{2(2n-3)!!\,\alpha_n}}
\right),
\label{eq:m_dynamics}
\end{align}
where $Dz:=\frac{dz}{\sqrt{2\pi}}e^{-z^2/2}$.
Here, $R_0$ represents the strength of the cross-talk noise. Since the signal term scales with $n$, we define the normalized parameter $\alpha_n^\prime=\alpha_n(2n-3)!!$,
and refer to it as the {\it loading rate}.
In the following, we use $\alpha_n^\prime$.

\subsection{Interpretation and correspondence with Amari--Maginu theory}

In general, in symmetric Hopfield-type models, feedback through recurrent couplings generates a retarded self-interaction in the effective single-unit dynamics, so that the current effective field depends explicitly on past neuron states.
The derived expression \eqref{eq:m_dynamics} has the same form as the overlap evolution equation obtained by neglecting the retarded self-interaction term in the modern Hopfield model \cite{mimura2025dynamical}.
Such a treatment corresponds to the Amari--Maginu signal-to-noise framework.
Although neglecting retarded self-interaction is an approximation in symmetric modern Hopfield models, this is not the case in the present sequential retrieval model.
Instead, the absence of retarded self-interaction follows naturally from the dynamical structure of the model, rather than being imposed as an assumption.
Consequently, Eq.~\eqref{eq:m_dynamics} is obtained without approximation, and the resulting overlap recursion coincides with the one obtained in the Amari–Maginu framework.

Motivated by the above discussion, we now show that Eq.~\eqref{eq:m_dynamics} can also be derived within the Amari--Maginu signal-to-noise framework.
For $\theta_i^{(t)}=0$ case, 
we decompose the local field \eqref{eq:local_field_asymmetric} into a signal term and a crosstalk noise term as
\begin{align}
    h_i^{(t)} = \xi_i^{t+1}n(m^{(t)})^{n-1}+\eta_i^{(t)}.
\end{align}
 The crosstalk noise is defined by
\begin{align}
    \eta_i^{(t)}=\sum_{\mu\neq t}^M 
	\xi_i^{\mu+1} n \biggl(m_\mu^{(t)}\biggr)^{n-1}.
\end{align}
The noise term satisfies $\mathbb E_{\bm{\xi}}[\eta_i^{(t)}]=0$, and $\mathbb E_{\bm{\xi}}[\eta_i^{(t)}\eta_j^{(t)}]=0$ for $i\neq j$.

Following the Amari–Maginu signal-to-noise approximation, we approximate the crosstalk noise as a temporally uncorrelated Gaussian variable. 
The contributions from different non-condensed patterns are uncorrelated to leading order, so the equal-time variance reduces to
\begin{align}
\mathbb E_{\bm{\xi}}\biggr[(\eta_i^{(t)})^2\biggr]
=n^2\mathbb E_{\bm{\xi}}\left[\sum_{\mu=1,\mu\neq t}(m_\mu^{(t)})^{2(n-1)}\right].
\end{align}

Assuming that $m_\mu^{(t)}$ is a zero-mean Gaussian random variable, its even moments are given by Wick's theorem (Isserlis's theorem),
\begin{align}
\mathbb E_{\bm{\xi}}[(m_\mu^{(t)})^{2(n-1)}]
=
(2n-3)!!
\left(\mathbb{V}[m_\mu^{(t)}]\right)^{n-1}.
\end{align}
Here, $\mathbb{V}[m_\mu^{(t)}]$ denotes the variance of $m_\mu^{(t)}$. Evaluating this variance within the same signal-to-noise approximation, we obtain
\begin{align}
\mathbb V[m_\mu^{(t)}]
&=
\frac1{N^2}\sum_{i,j}
\mathbb E_{\boldsymbol\xi}
[\xi_i^\mu\xi_j^\mu S_i^{(t)}S_j^{(t)}] \\
&=
\frac1{N^2}\sum_i(S_i^{(t)})^2
=\frac1N,
\end{align}
for any $\mu$ and $t$. Hence
\begin{align}
    \mathbb E_{\bm{\xi}}\biggr[\biggl(\eta_i^{(t)}\biggr)^2\biggr]=n^2(2n-3)!!\alpha_n,
\end{align}
which is equal to the variance of effective Gaussian in \eqref{eq:m_final}.

The Gaussian signal-to-noise analysis reproduces the exact one-time
overlap dynamics obtained by GFA. Although the effective Gaussian noise obtained from GFA may possess
temporal correlations, these correlations do not enter the one-time recursion for the overlap, which depends
only on the equal-time variance $R^{(t,t)}$. 
Thus, the Amari--Maginu-type scalar recursion becomes exact for the
overlap dynamics in the present sequential retrieval model.

\section{Numerical Validation}
\label{sec:validation}

We present the results obtained from the GFA and compare them with direct numerical simulations of \eqref{eq:def_neuron_update} using the local field defined in \eqref{eq:local_field_asymmetric}.
When the interaction order is fixed or clear from context, we abbreviate \(\alpha'_n\) as \(\alpha'\). In the following figures, we set \(n=3\), so that \(\alpha'=\alpha'_3\).

\subsection{Recalling trajectories}
\label{sec:simulation}

\begin{figure*}[htbp]
  \begin{minipage}{.48\linewidth}
    \centering
    \includegraphics[width=\linewidth]{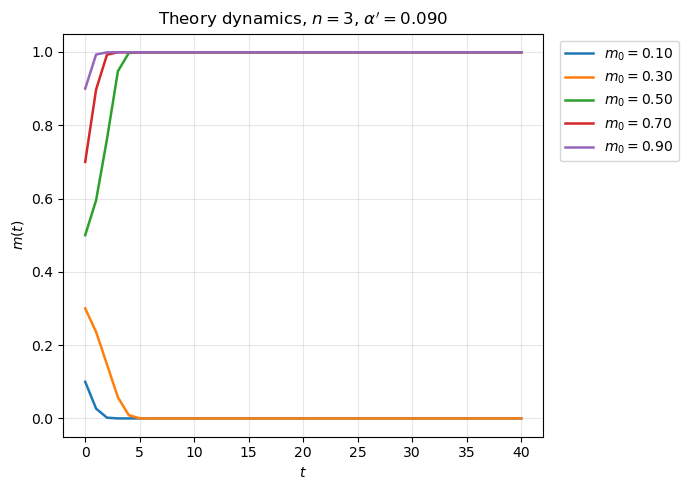}
  \end{minipage}
  \begin{minipage}{.48\linewidth}
    \centering
    \includegraphics[width=\linewidth]{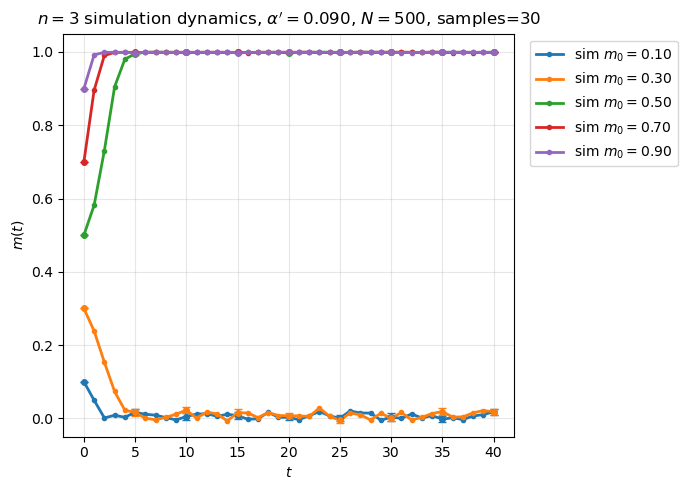}
  \end{minipage}
  \caption{Sequential retrieval process at $\alpha'=0.09$.
Left: results obtained from GFA. 
Right: results from direct numerical simulations for $N=500$ neurons, averaged over 30 trials. Error bars indicate the standard error and are shown every five iteration steps for visual clarity. The horizontal and vertical axes represent the iteration step and the overlap, respectively. }
  \label{fig:alpha0.09}

\end{figure*}

\begin{figure*}[htbp]
  \begin{minipage}{.48\linewidth}
    \centering
    \includegraphics[width=\linewidth]{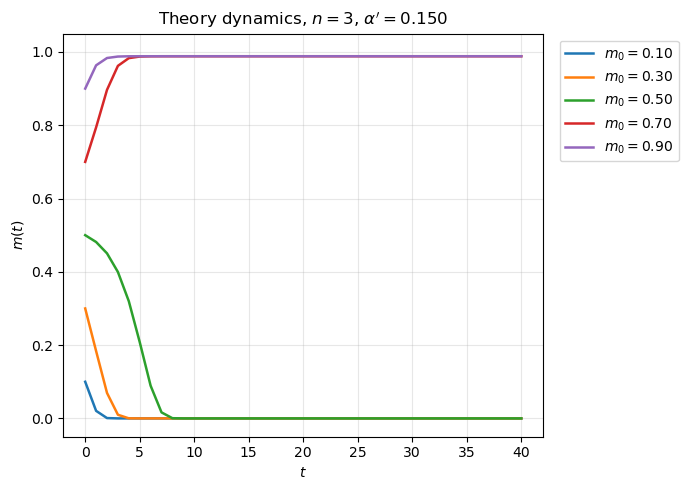}
  \end{minipage}
  \begin{minipage}{.48\linewidth}
    \centering
    \includegraphics[width=\linewidth]{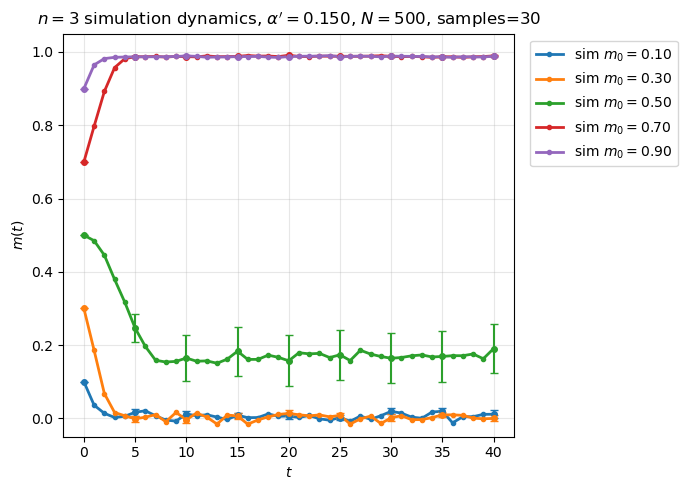}
  \end{minipage}
\caption{Sequential retrieval process at $\alpha'=0.15$.
Left: results obtained from the GFA.
Right: results from direct numerical simulations.
The simulation conditions and axis definitions are the same as those in Fig.~\ref{fig:alpha0.09}.}
\label{fig:alpha0.15}
  
\end{figure*} 

\begin{figure*}[htbp]
  \begin{minipage}{.48\linewidth}
    \centering
    \includegraphics[width=\linewidth]{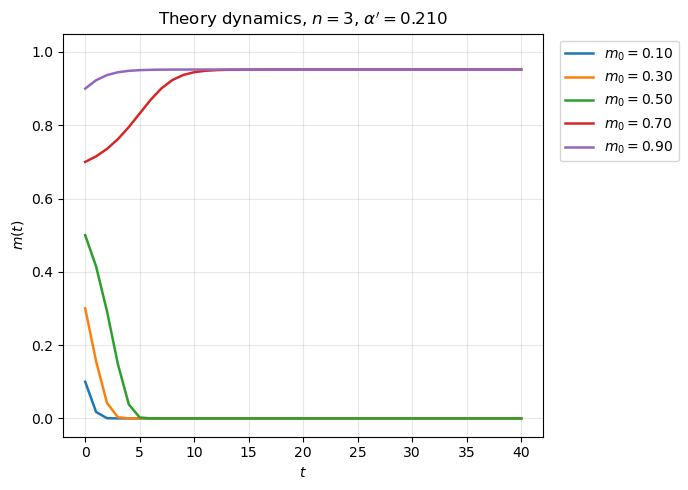}
  \end{minipage}
  \begin{minipage}{.48\linewidth}
    \centering
    \includegraphics[width=\linewidth]{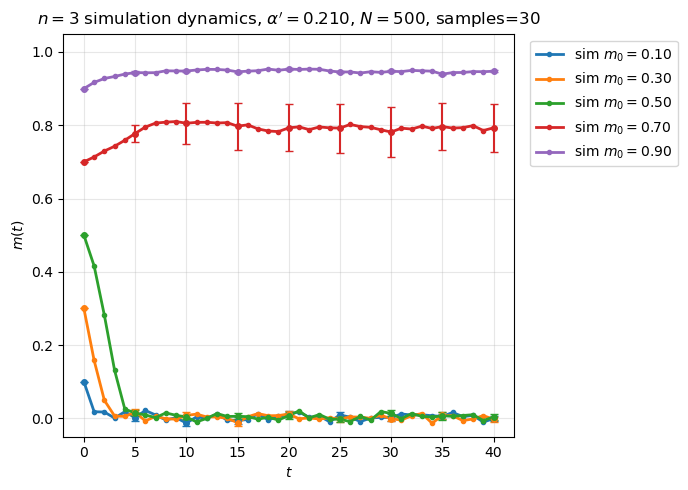}
  \end{minipage}
\caption{Sequential retrieval process at $\alpha'=0.21$.
Left: results obtained from the GFA.
Right: results from direct numerical simulations.
The simulation conditions and axis definitions are the same as those in Figs.~\ref{fig:alpha0.09}-\ref{fig:alpha0.15}.}
\label{fig:alpha0.21}
\end{figure*}

\begin{figure*}[htbp]
  \begin{minipage}{.48\linewidth}
    \centering
    \includegraphics[width=\linewidth]{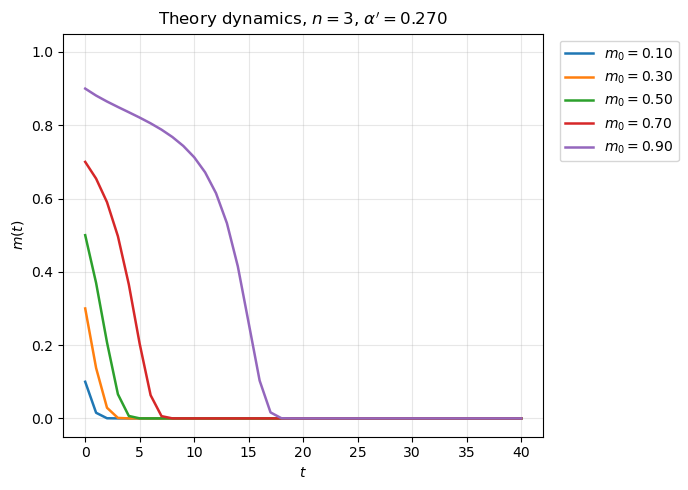}
  \end{minipage}
  \begin{minipage}{.48\linewidth}
    \centering
    \includegraphics[width=\linewidth]{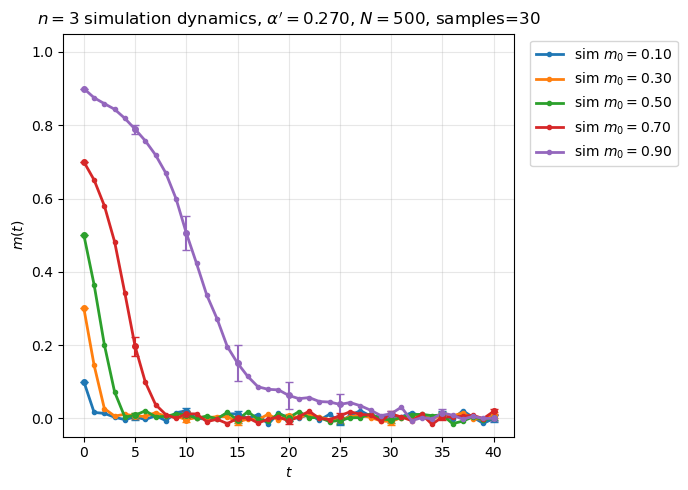}
  \end{minipage}
  \caption{Sequential retrieval process at $\alpha'=0.27$.
Left: results obtained from the GFA.
Right: results from direct numerical simulations.
The simulation conditions and axis definitions are the same as those in Figs.~\ref{fig:alpha0.09}-\ref{fig:alpha0.21}.}
  \label{fig:alpha0.27}

\end{figure*}

Figs.~\ref{fig:alpha0.09}--\ref{fig:alpha0.27} show the time evolution of the overlap $m^{(t)}$ for $n=3$ with multiple initial overlaps at $\alpha^\prime=0.09$ (Fig.~\ref{fig:alpha0.09}), $\alpha^\prime=0.15$ (Fig.~\ref{fig:alpha0.15}), $\alpha^\prime=0.21$ (Fig.~\ref{fig:alpha0.21}), and $\alpha^\prime=0.27$ (Fig.~\ref{fig:alpha0.27}). In each figure, the left panel shows the numerical solution of the GFA dynamical equations \eqref{eq:m_time_evolution}, while the right panel shows the results of direct numerical simulations of the network with $N=500$ neurons, averaged over 30 independent trials.
As the loading parameter $\alpha^\prime$ increases, the minimum initial overlap required for convergence to the nonzero retrieval fixed point \(m_s(\alpha')\) also increases.
 In other words, an increasingly larger initial overlap is required for the trajectory to converge to nonzero fixed point, whereas trajectories starting from smaller initial overlaps fail to do so. This dependence on the initial condition is in good agreement between the GFA and the direct numerical simulations.

Quantitative differences can nevertheless be observed in the transient dynamics and, in some cases, in the final converged value of the overlap. 
These discrepancies are particularly pronounced for intermediate initial overlaps. 
Similar discrepancies between theoretical predictions and numerical simulations have also been observed in the retrieval dynamics of Modern Hopfield models \cite{mimura2025dynamical}. 
These differences are likely attributable to finite-size effects in the numerical simulations.

Overall, the GFA successfully captures the essential dynamical behavior of the sequential retrieval process and shows good agreement with direct numerical simulations.

\subsection{Basin structure: Retrieval possible region}

The dependence of the retrieval dynamics on the initial condition discussed above can be summarized by analyzing the region in the $(\alpha',m^{(0)})$ plane where the dynamics converge to a retrieval state ($m>0$).
This analysis quantifies how the loading parameter and the initial overlap affect the retrieval behavior.
The storage capacity is defined as the critical loading parameter above which no initial overlap leads to a retrieval state and denoted by $\alpha^\prime_{c,n}$. When \(n\) is fixed, we write it simply as $\alpha^\prime_c$.

\begin{figure}
    \centering
    \includegraphics[width=0.9\linewidth]{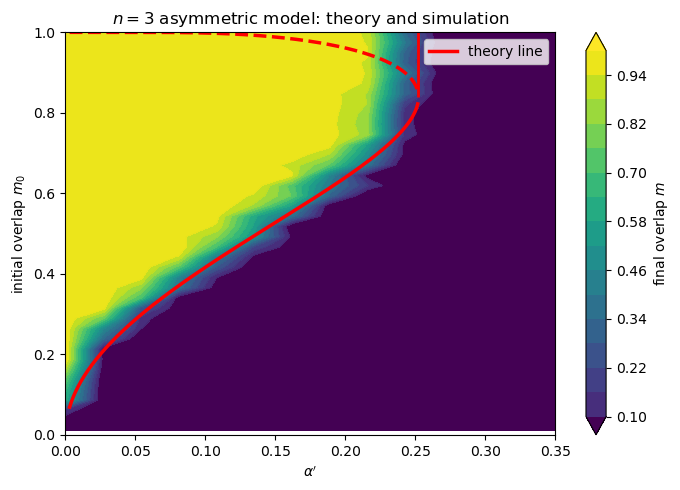}
    \caption{Basin of attraction to the retrieval state ($m>0$) for $n=3$. The color map shows the results of numerical simulations for $N=200$, where the overlap after 80 steps, starting from an initial overlap $m^{(0)}$, is averaged over 30 trials. The solid line represents the theoretical prediction, and the dashed line represents the stable fixed point.}
    \label{fig:basin_n3}
\end{figure}

Figure~\ref{fig:basin_n3} shows the basin of attraction for $n=3$, where the vertical and horizontal axes represent the initial overlap $m^{(0)}$
and the loading rate $\alpha^\prime$, respectively. 
The color map shows the overlap after $80$ steps of the update, starting from $m^{(0)}$ at $\alpha^\prime$, numerically obtained at $N=200$ and averaged over $30$ trials. 
The basin boundary -- the smallest $m^{(0)}$
above which the dynamics converges to $m>0$ (equivalently, below which it converges to $m=0$) -- derived from theory, is shown by the solid line.
These results show that the theoretical macroscopic dynamics captures the retrieval behavior of the model with good accuracy. 
The threshold, i.e. the critical normalized loading rate for $n=3$ is $\alpha'_{c,3}=0.252$. For comparison, the pairwise sequence-processing model has a reported storage capacity of $0.269$ \cite{during1998phase}. This value is obtained from the \(n=2\) theory, in which response-dependent contributions enter the crosstalk covariance; the scalar recursion derived here applies only for $n\ge3$.

The self-consistent equation \eqref{eq:m_final} for the macroscopic overlap admits three fixed points simultaneously for $0<\alpha'<\alpha'_c$: $m=0$, an unstable fixed point $0<m_u<m_s$, and a stable retrieval fixed point $m_s$. At $\alpha'=\alpha'_c$, the two finite-overlap fixed points merge, $m_u=m_s=m_c$, and for $\alpha'>\alpha'_c$, only $m=0$ remains.
The $\alpha^\prime$-dependence of the stable fixed point is shown by the dashed line in Fig.\ref{fig:basin_n3}.

For \(n\ge3\),
denoting the right-hand side of Eq.~\eqref{eq:m_dynamics} by f(m), one has
one has \(f'(0)=0\), so \(m=0\) is locally stable for every \(\alpha'>0\). For \(0<\alpha'<\alpha'_c\), this stable zero-overlap fixed point coexists with a stable retrieval fixed point \(m_s\), separated by the unstable fixed point \(m_u\). Therefore, it acts as the basin boundary: 
trajectories starting from $m^{(0)}<m_u$ flow to $m=0$, while those from $m^{(0)}>m_u$ flow to $m_s$. 
Hence the solid line (basin boundary) coincides with the unstable fixed point.
As $\alpha^\prime$ increases toward $\alpha_c^\prime$, the stable fixed point $m_s$ and the unstable fixed point $m_u$ approach each other and merge in a saddle-node bifurcation at $\alpha^\prime=\alpha_c^\prime$, disappearing simultaneously and leaving $m=0$ as the only fixed point for $\alpha'>\alpha_c^\prime$. 
Since both branches vanish together at this point, the basin boundary line becomes vertical at $\alpha'=\alpha_c^\prime$.

\begin{figure}
    \centering
    \includegraphics[width=\columnwidth]{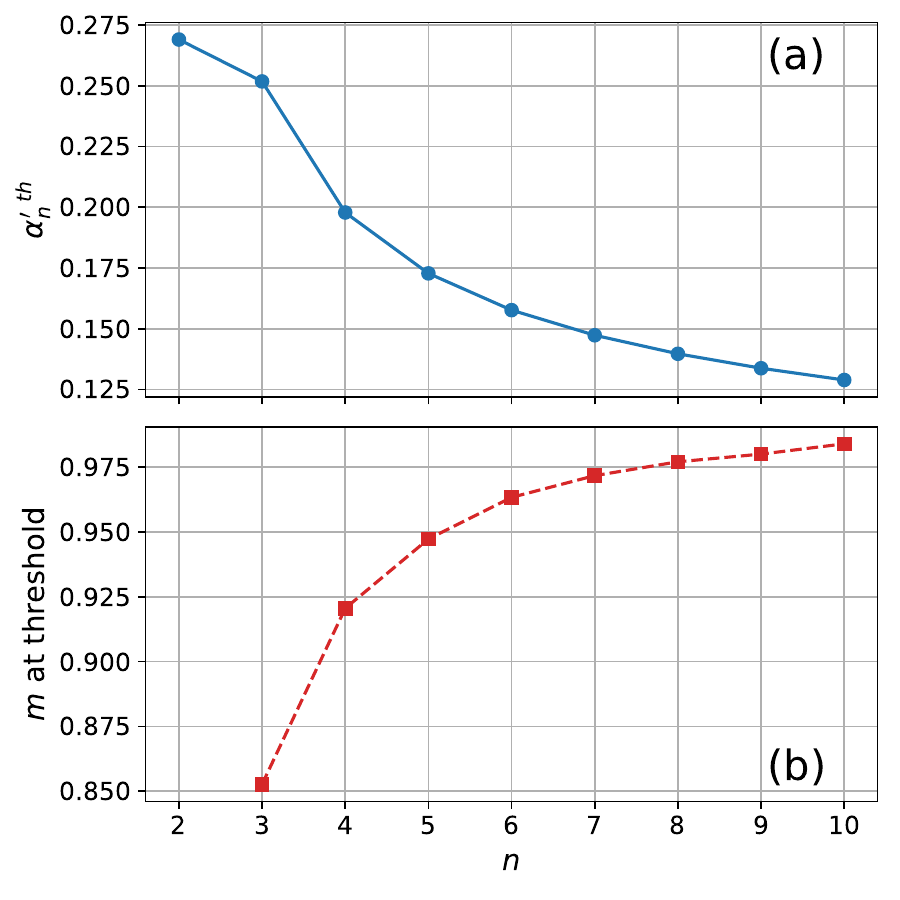}
    \caption{$n$-dependence of (a) the threshold $\alpha_{c,n}^\prime$ and (b) the critical overlap $m_c$.  
    }
    \label{fig:threshold_vs_n}
\end{figure}

Fig.\ref{fig:threshold_vs_n} (a) shows the $n$-dependence of the threshold value of $\alpha_c^\prime$.
As $n$ increases, the threshold monotonically decreases.
In addition, Fig.~\ref{fig:threshold_vs_n}(b) shows the value of the fixed point $m$ at the threshold, which increases with $n$. This indicates that, for larger $n$, the transition to the stable solution at $m=0$ becomes more abrupt. 
This behavior can be attributed to the effect of the power $n$ in the macroscopic dynamics \eqref{eq:m_dynamics}: as $n$ increases, $m^{n-1}$ decreases increasingly rapidly as $m$ falls below unity, leading to a sharper transition toward the stable solution at $m=0$.

We next consider the large-$n$ behavior of the retrieval threshold. At the critical loading, the stable and unstable finite-overlap fixed points merge in a saddle-node bifurcation. From \eqref{eq:m_dynamics}, the critical loading $\alpha'_{c,n}$ and overlap $m_c$ are therefore determined by solving the following equations: 
\begin{align}
    m&=
    \operatorname{erf}\left(\frac{m^{n-1}}{\sqrt{2 \alpha'_n}}\right),
    \label{eq:crit_m}
    \\
    1&=
    \frac{d}{dm}
    \operatorname{erf}\left(\frac{m^{n-1}}{\sqrt{2 \alpha'_n}}\right). 
    \label{eq:crit_deriv}
\end{align}
Using the asymptotic expansion of the error function $\operatorname{erf}(z)=1-(\sqrt{\pi}z)^{-1} e^{-z^2} [1+O(z^{-2})]$ for a large $z$, we obtain
\begin{align}
    \alpha'_{c,n} &=
    \frac{1}{2\ln n}\left[1+O\left(\frac{\ln\ln n}{\ln n}\right)\right],
    \label{eq:large_n_alpha}\\
    m_c&=
    1-\frac{1}{2n\ln n}\left[1+O\left(\frac{\ln\ln n}{\ln n}\right)\right].
    \label{eq:large_n_mc}
\end{align}
The derivation can be found in Appendix \ref{sec:app_abs_c}. Thus, the normalized critical loading decreases as $1/\ln n$, while the critical overlap approaches unity. 
\par
Since $M=\alpha_n N^{n-1}$ and $\alpha'_n=(2n-3)!!\,\alpha_n$, the large-$n$ retrieval capacity behaves as 
\begin{align}
    M_c^{\mathrm{ret}} \sim \frac{N^{n-1}}{2(2n-3)!!\,\ln n}.
    \label{eq:large_n_retrieval_capacity}
\end{align}
For comparison, Chaudhry \textit{et al.} derived the capacity for perfect single-step transitions of the same model, in which the next pattern is reproduced perfectly when the network starts from a stored pattern \cite{chaudhry2023longsequencehopfieldmemory}.
It coincides with absolute-capacity estimate of the single-site criterion by Krotov and Hopfield \cite{krotov2016dense} for unbiased polynomial dense associative memory.
\begin{align}
    M_c^{\mathrm{abs}} \sim \frac{N^{n-1}}{2(2n-3)!!\,\ln N}.
    \label{eq:absolute_capacity_comparison}
\end{align}
These expressions differ only in the logarithmic factor. In the case where the interaction order grows sublinearly with the system size as $n=N^\gamma$ with $0<\gamma<1$, \eqref{eq:large_n_retrieval_capacity} and \eqref{eq:absolute_capacity_comparison} give 
\begin{align}
    M_c^{\mathrm{ret}}
    \sim \frac{1}{\gamma} M_c^{\mathrm{abs}}. 
\end{align}
The present theory, however, takes the thermodynamic limit at fixed interaction order. Establishing this relation in a joint large-$N$, large-$n$ limit requires a uniform asymptotic analysis of the Gaussian moment estimates underlying the factor $(2n-3)!!$.

\section{Discussion and Conclusion}
\label{sec:conclusion}

In this paper, we have studied an asymmetric modern Hopfield model that exhibits sequential retrieval of patterns.
We have applied the GFA and characterized the limit cycle solution representing sequential retrieval using macroscopic parameters. 
The derived macroscopic dynamics of retrieval is expressed in terms of the overlaps with embedded patterns and Gaussian noise with a time-independent equal-time variance, matching the form obtained via signal-to-noise analysis under the Amari-Maginu approximation. 
Our analytical expression successfully captured the dynamics consistent with numerical simulations, in particular, the dependence of retrieval dynamics on initial conditions and corresponding basin volume aligned well with the numerical results.

We have found that in the modern sequential processing network, the normalized critical loading rate monotonically decreases as the order of interaction increases.  
This stands in contrast to the conventional symmetric modern Hopfield model, where increasing $n$ from 2 to 3 enhances the storage capacity. The comparison suggests that the absence of a retarded self-interaction may play an important role, although establishing this connection quantitatively requires a separate analysis of the symmetric model.

A natural extension is to investigate retrieval dynamics for finite sequence lengths. In pairwise sequence-processing models, finite cycles give rise to retarded self-interactions \cite{mimura2004path}. Comparing pairwise and higher-order models at the same cycle lengths may help clarify how the interaction order affects retrieval dynamics and relaxation times.

Our theoretical framework offers a physics-based perspective on temporal sequence generation in modern AI architectures. 
Standard modern Hopfield models mathematically correspond to spatial attention mechanisms, hence introducing interactions constructed from shifted patterns naturally drives time-dependent state transitions.
Although such sequential processing naturally lies outside thermal equilibrium, our study has demonstrated that statistical-mechanical tools remain remarkably powerful in characterizing these driven dynamics. 
This work may provide a useful theoretical framework for analyzing time-dependent processes at the intersection of machine learning and physics.

\section*{Acknowledgments}
The authors (MM and AS) thank Koji Hukushima, Yoshiyuki Kabashima, and Takashi Takahashi for supporting this research by providing space and facilities. This work was supported by the RIKEN-AIP Undergraduate Research Program (MM), JST PRESTO JPMJPR23J4 (AS) and JSPS KAKENHI Grant Nos. JP22H05117 (AS), JP23K03841, JP23H05492, JP25K24611, and JP26K02981 (KM).

\appendix 
\section{Detailed sketch of proposition 1}
\label{sec:app_prop_A}

The disorder average is evaluated following the calculation for symmetric dense associative memory in Appendix A of Ref.~\cite{mimura2025dynamical}. We use the same macroscopic order parameters and combinatorial organization of the pattern averages. The common steps are included for completeness, while the modifications arising from the shifted output-pattern index, most notably the vanishing of cross terms between different patterns, Eq.~\eqref{eq:app_covariance_0} are treated explicitly below.

Here we evaluate \eqref{eq:Xi}.
Introducing the relationships \eqref{eq:def_mt} and \eqref{eq:def_k}
through delta-function constraints, we obtain
\begin{align}
\nonumber
&\Xi(\bm{S}^{0:T})=\int d\bm{m}d\bm{k}
    \exp\biggl(-inN\sum_{t=0}^{T-1}k^{t+1,t}\left(m^{(t)}\right)^{n-1}\biggr)\\
    &\times\mathbb{E}_{\{\bm{\xi}^{\mu}\}}
\biggl[\exp\biggl(
-i\sum_{\mu=1}^MX_\mu
\biggr)
\prod_{t=0}^{T-1}
\delta\biggl(
m^{(t)}-\frac1N\sum_{i=1}^{N}\xi_i^t S_i^{(t)}
\biggr)\nonumber\\
&\times
\prod_{t=0}^{T-1}\delta\biggr(k^{t+1,t}-\frac{1}{N}\sum_{i=1}^N\xi_i^{t+1}\hat{h}_i^{(t)}\biggl)
\biggr].
\label{eq:app_Xi_1}
\end{align}
To perform the average over the random patterns,
we first examine the statistical properties of $X_\mu$.
Since $X_\mu$ depends linearly on
$\bm\xi^{\mu+1}$ and the remaining factor is independent of
$\bm\xi^{\mu+1}$, we have
\[
\mathbb{E}[X_\mu]=0 .
\]
Furthermore, from the definition \eqref{eq:def_X_mu},
\begin{align}
\nonumber
    X_\mu X_\nu&=n^2\!\!\!\sum_{t=0(\neq\mu)}^{T-1}
        \sum_{t^\prime=0(\neq\nu)}^{T-1}\sum_{i=1}^{N}\sum_{j=1}^N\hat h_i^{(t)}\hat{h}_j^{(t^\prime)}
\xi_i^{\mu+1}\xi_j^{\nu+1}\\
&\times
\biggr(\frac1N\sum_{k\ne i}^{N}\xi_k^{\mu}S_k^{(t)}\biggr)^{n-1} 
\biggr(\frac1N\sum_{\ell\ne j}^{N}\xi_\ell^{\nu}S_\ell^{(t^\prime)}\biggr)^{n-1}.
\label{eq:app_X_2}
\end{align}
For $\mu\neq\nu$, the two terms may share a pattern only when $\mu$ and $\nu$ are neighboring indices in the stored sequence. Even in that case, at least one of the shifted output-pattern vectors $\xi^{\mu+1}$ and $\xi^{\nu+1}$ is independent of all the remaining factors and enters linearly. 
Averaging over this unbiased and independent pattern therefore gives
\begin{align}
    \mathbb{E}[X_\mu X_\nu] = 0\quad (\mu\neq\nu),
    \label{eq:app_covariance_0}
\end{align}
so that cross terms involving different pattern indices do not contribute.
In addition, $\mathbb{E}[X_\mu^3]\sim O(n^3TN^{-3(n-2)/2})$ for any $\mu$, 
and the higher-order cumulants are of even smaller order and can be neglected for $n\ge3$.
Therefore, the Taylor expansion with respect to $X_\mu$ can be truncated at second order.
Hence we can approximate \eqref{eq:app_Xi_1} as 
\begin{align}
\nonumber
&\Xi(\bm{S}^{0:T})=\int d\bm{m}d\bm{k}
    \exp\biggl(-inN\sum_{t=0}^{T-1}k^{t+1,t}\left(m^{(t)}\right)^{n-1}\biggr)\\
    &\times\mathbb{E}_{\{\bm{\xi}^{\mu}\}}
\biggl[\prod_{\mu=1}^M\biggl(1-\frac{1}{2}X_\mu^2
\biggr)
\prod_{t=0}^{T-1}
\delta\biggl(
m^{(t)}-\frac1N\sum_{i=1}^{N}\xi_i^t S_i^{(t)}
\biggr)\nonumber\\
&\times
\prod_{t=0}^{T-1}\delta\biggr(k^{t+1,t}-\frac{1}{N}\sum_{i=1}^N\xi_i^{t+1}\hat{h}_i^{(t)}\biggl)
\biggr].
\label{eq:app_Xi_temp}
\end{align}
The property \eqref{eq:app_covariance_0} is characteristic of the sequential associative memory model for any $n(\geq 2)$ whereas $\mathbb{E}[X_\mu X_\nu]~(\mu\neq\nu)$ have finite values in the symmetric case, as in \cite{mimura2025dynamical}. For $n\ge3$, this further leads to the reduced form \eqref{eq:app_Xi_temp}.

Now we compute $\mathbb{E}_{\bm{\xi}}[X_\mu^2]$.
Following \eqref{eq:app_X_2} with $\mu=\nu$, and expanding the power
$n-1$, we obtain
\begin{align}
        \mathbb{E}_{\{\bm{\xi}\}}&[X_\mu^2]=\frac{n^2}{N^{2(n-1)}}\!\!\!\sum_{t=0(t\neq\mu)}^{T-1}\sum_{t^\prime=0(t^\prime\neq\mu)}^{T-1}\sum_{i=1}^{N}\sum_{j=1}^N\hat h_i^{(t)}\hat{h}_j^{(t^\prime)}\mathcal{N}_{ij}^{\mu,t,t^\prime},
        \label{eq:app_X_variance}
\end{align}
where we set 
\begin{align}
\nonumber
	\mathcal{N}_{ij}^{\mu,t,t^\prime}
	=& \mathbb{E}_{\{\bm{\xi}\}}\left[\xi_i^{\mu+1}\xi_j^{\mu+1}\biggr(\sum_{\kappa\ne i}^{N}\sum_{\kappa^\prime\ne j}^{N}\xi_\kappa^{\mu}S_\kappa^{(t)}\xi_{\kappa^\prime}^{\mu}S_{\kappa^\prime}^{(t^\prime)}\biggr)^{n-1}\right]\\
	=&\mathbb{E}_{\{\bm{\xi}\}}\biggl[
	\xi_i^{\mu+1} \xi_{j} ^{\mu+1}
	\sum_{{\kappa}_1  \ne i }^N \cdots\!\!\! \sum_{\kappa_{n-1}  \ne i }^N 
	\sum_{\kappa^\prime_1 \ne j}^N \cdots\!\!\! \sum_{\kappa^\prime_{n-1} \ne j}^N \notag\\
	&\xi_{\kappa_1 }^{\mu} \cdots \xi_{\kappa_{n-1}}^{\mu}  
	\xi_{\kappa^\prime_1}^{\mu} \cdots \xi_{\kappa^\prime_{n-1}} ^{\mu}
	S_{\kappa_1 }^{(t )} \cdots S_{\kappa_{n-1} }^{(t )} 
	S_{\kappa^\prime_1}^{(t')} \cdots S_{\kappa^\prime_{n-1}}^{(t')} 
	\biggr]. 
    \label{eq:app_N_def}
\end{align}
The quantity $\mathcal{N}_{ij}^{\mu,t,t^\prime}$ is nonzero only when $i=j$ and the
indices $\kappa_1,\ldots,\kappa_{n-1}$ and $\kappa^\prime_1,\ldots,\kappa^\prime_{n-1}$
can be arranged into $(n-1)$ pairs.
In other words, every pattern
variable must be paired with an identical pattern variable for the
pattern products to survive the average over the random patterns.

There are three types of pairings: pairs between two ``unprimed" indices $\kappa$s,
pairs between two ``primed" indices $\kappa^\prime$s, and pairs between one unprimed and one
primed index. 
We denote by $A(n-1,k)$ the number of
pairings in which exactly $k$ pairs connect one unprimed index and one primed index. 
This number is given by \eqref{eq:def_A}, and \(B(m)\), defined by \eqref{eq:def_B} is the number of pairings among \(m\) indices of the same type. The identity
\[
\sum_{k=0}^{n-1} A(n-1,k)=B(2(n-1))
\]
is also useful.

The leading-order term of \eqref{eq:app_N_def} for $i=j$ can be expressed using $A(\cdot)$, and is of order $O(N^{n-1})$, while the remaining terms are $O(N^{n-2})$ or smaller. Since the denominator of \eqref{eq:app_X_variance} contains $N^{2(n-1)}$, it is sufficient to consider only the leading-order term of \eqref{eq:app_N_def} .
We therefore truncate \eqref{eq:app_N_def} at the leading order as follows:
\begin{align}
\mathcal{N}_{ij}^{\mu,t,t^\prime}
\!\!\!=
\begin{cases}
N^{n-1}\!
\displaystyle\sum_{k=0}^{n-1}
\! A(n-1,k)
\!\left(
q^{(t,t^\prime)}
\right)^k
&
(i=j),
\\[2mm]
0,
&
(i\neq j),
\end{cases}
\label{eq:app_calN}
\end{align}
where we use \eqref{eq:def_q}. 

The terms in \eqref{eq:app_calN} are independent of $(i,j)$ and $\mu$. Using this property and denoting $\mathcal{N}_{ii}^{\mu,t,t^\prime}=N^{n-1}W^{t,t^\prime}$,
we obtain
\begin{align}
        \mathbb{E}_{\{\bm{\xi}\}}&[X_\mu^2]=\frac{n^2}{N^{n-2}}\sum_{t=0(t\neq\mu)}^{T-1}\sum_{t^\prime=0(t^\prime\neq\mu)}^{T-1}Q^{t,t^\prime}W^{t,t^\prime},
\end{align}
where we use \eqref{eq:def_Q}.
Therefore,
\begin{align}
\nonumber
&\Xi(\bm{S}^{0:T})\!=\!\!\!\int \!\!d\bm{m}d\bm{k}d\underline{q}d\underline{Q}d\underline{K}
    \exp\biggl(-inN\!\sum_{t=0}^{T-1}k^{t+1,t}\left(m^{(t)}\right)^{n-1}\biggr)\\
    &\times\prod_{\mu=1}^M\biggl(1-\frac{1}{2}\frac{n^2}{N^{n-2}}\sum_{t=0(t\neq\mu)}^{T-1}\sum_{t^\prime=0(t^\prime\neq\mu)}^{T-1}Q^{t,t^\prime}W^{t,t^\prime}
\biggr)\notag \\
&\times {\cal W}\left(\bm{m},\bm{k},\underline{q},\underline{Q},\underline{K};\bm{S}^{0:T},\widehat{\bm{h}}\right),
\end{align}
where ${\cal W}\left(\bm{m},\bm{k},\underline{q},\underline{Q},\underline{K};\bm{S}^{0:T},\widehat{\bm{h}}\right)$ counts the number of microscopic states $\{\bm{\xi}^\mu\}$ that give the specified values of $(\bm{m},\bm{k},\underline{q},\underline{Q},\underline{K})$, for given $(\bm{S}^{0:T},\widehat{\bm{h}})$, and is defined by
\begin{align}
    &{\cal W}\left(\bm{m},\bm{k},\underline{q},\underline{Q},\underline{K};\bm{S}^{0:T},\widehat{\bm{h}}\right) \notag\\
    &=\mathbb{E}_{\{\bm{\xi}^{\mu}\}}
\biggl[\prod_{t=0}^{T-1}
\delta\biggl(
m^{(t)}-\frac1N\sum_{i=1}^{N}\xi_i^t S_i^{(t)}
\biggr)\nonumber\\
&\hspace{1.0cm}\times
\prod_{t=0}^{T-1}\delta\biggr(k^{t+1,t}-\frac{1}{N}\sum_{i=1}^N\xi_i^{t+1}\hat{h}_i^{(t)}\biggl) \notag\\
&\hspace{1.0cm}\times \prod_{t,t^\prime}\delta\biggr(q^{(t,t^\prime)}-\frac1N\sum_{i=1}^{N}S_i^{(t)} S_i^{(t')}\biggl) \notag\\
&\hspace{1.0cm}\times \prod_{t,t^\prime}\delta\biggr(Q^{(t,t^\prime)}-\frac1N\sum_{i=1}^{N}\hat h_i^{(t)} \hat h_i^{(t')}\biggl) \notag\\
&\hspace{1.0cm}\times
\prod_{t,t^\prime}\delta\biggl(K^{(t,t')}-
\frac1N\sum_{i=1}^{N}S_i^{(t)}\widehat h_i^{(t')}\biggr)
\biggr].
\end{align}
In summary, for $N\to \infty$,
\begin{align}
\nonumber
&\Xi(\bm{S}^{0:T})=\int d\bm{m}d\bm{k}d\underline{q}d\underline{Q}d\underline{K}
    \exp\biggl\{-inN \sum_{t=0}^{T-1}k^{t+1,t}\left(m^{(t)}\right)^{n-1}\biggr\}\\
    &\times\exp\biggl(-\frac{n^2M}{2N^{n-2}}\sum_{t,t^\prime}^{T-1}Q^{t,t^\prime}W^{t,t^\prime}
\biggr)\notag \\
&\times {\cal W} \left(\bm{m},\bm{k},\underline{q},\underline{Q},\underline{K};\bm{S},\widehat{\bm{h}}\right),
\label{eq:app_Xi_final}
\end{align}
where we used
\begin{equation}
\sum_{\mu=1}^{M}
\mathbf 1_{\mu\ne t}\mathbf 1_{\mu\ne t'}
=
M+O(1).
\end{equation}

We now compute ${\cal W}$ explicitly.
Introducing integral expression of delta functions, we obtain
\begin{align}
&{\cal W}=\int d\hat{\bm{m}}d\hat{\bm{k}}d\hat{\underline{q}}d\hat{\underline{Q}}d\hat{\underline{K}} \notag\\
&\times\exp\biggl\{iN\sum_{t=0}^{T-1}m^{(t)}\hat{m}^{(t)}+iN\sum_{t=0}^{T-1}k^{(t+1,t)}\hat{k}^{(t+1,t)} \notag\\
&\hspace{0.5cm}+iN\sum_{t=0}^{T-1}\sum_{t^\prime=0}^{T-1}\biggl(
q^{(t,t')}\hat q^{(t,t')}
+
Q^{(t,t')}\hat Q^{(t,t')}
+
K^{(t,t')}\hat K^{(t,t')}\biggr)\biggr\}\notag \\
&\times\prod_{i=1}^N
\mathbb E_{\{\xi_i^\mu\}}
\biggl[\exp\biggl\{-i\sum_{t=0}^{T-1}\hat{m}^{(t)}\xi_i^tS_i^{(t)}-i\sum_{t=0}^{T-1}\hat{k}^{t+1,t}\xi_i^{t+1}\widehat{h}_i^{(t)}\biggr\}\biggr]\notag \\
&\times\prod_{i=1}^N\biggl[\exp\biggl\{-i\!\sum_{t=0}^{T-1}\!\sum_{t^\prime=0}^{T-1}\biggl(\hat{q}^{(t,t^\prime)}S_i^{(t)}S_i^{(t^\prime)}\notag\\
&\hspace{1.2cm}+\hat{Q}^{(t,t^\prime)}\widehat{h}_i^{(t)}\widehat{h}_i^{(t^\prime)}\!\! +\!
\hat K^{(t,t')}S_i^{(t)}\hat h_i^{(t')}\!\biggr)\!\biggr\}\!\biggr].
\label{eq:app_delta_functions}
\end{align}

Combining \eqref{eq:noise-in-Z}, \eqref{eq:app_Xi_final} and \eqref{eq:app_delta_functions}, we obtain Proposition \ref{prop:Z_form}.

\section{Detailed sketch of proposition 2}
\label{sec:app_prop_B}

The derivation follows the saddle-point analysis and the
construction of the effective single-unit process for symmetric dense associative memory presented in Appendix B of Ref.~\cite{mimura2025dynamical}. We include the common intermediate steps for completeness and explicitly identify the changes arising from the asymmetric interaction.
\par
Starting from the generating functional in Proposition~1 and the single-unit measure defined in Eq.~\eqref{eq:single_unit_measure}, we derive Eq.~\eqref{eq:def_average}. In the present model, the disorder contribution $\Omega$ is independent of $K$, which yields $\widehat{K}=0$ at the saddle point and eliminates the retarded self-interaction from the effective single-unit dynamics.
\color{black}

\subsection{Saddle Points of ${\cal Q}$}
\label{sec:app_saddle}

\paragraph{Two-time correlation of neurons $q^{(t,t^\prime)}$:}

From \eqref{eq:Z_ave_prop1}, we obtain the following relationships:
\begin{align}
	& 
	\lim_{\bm{\psi}\to\bm{0}} \!
	\frac{\partial^2 \bar{Z}[\bm{\psi}]}
	{\partial \psi_i^{(t)} \partial \psi_{j}^{(t')}} 
	\!=\! 
	-\delta_{i,j} 
	\langle S^{(t)} S^{(t')}\rangle_i 
	\!-\! (1\!-\!\delta_{i,j}) 
	\langle S^{(t)}\rangle_i 
	\langle S^{(t')}\rangle_{j},
\end{align}
where $\delta_{i,j}$ is the Kronecker's delta, and $\bm{\psi}\to \bm{0}$ denotes the limit in which all components $\psi_j^{(t)}$ are set to zero.
Therefore, at the saddle point, we obtain
\begin{align}
    q^{(t,t^\prime)}=-\frac{1}{N}\sum_{i,j}\delta_{i,j}\lim_{\bm{\psi}\to\bm{0}} \!
	\frac{\partial^2 \overline{Z}[\bm{\psi}]}
	{\partial \psi_i^{(t)} \partial \psi_{j}^{(t')}}.
    \label{eq:app_Z_derivative_psi2}
\end{align}

Meanwhile, from the definition of 
$\overline{Z}[\bm{\psi}]$ (eq.\eqref{eq:def-Z}), we obtain
\begin{align}
    \lim_{\bm{\psi}\to\bm{0}}\frac{\partial^2\overline{Z}[\bm{\psi}]}{\partial\psi_i^{(t)}\partial\psi_j^{(t^\prime)}}&=-\mathbb{E}_{\{\bm{\xi}^\mu\}}\biggl[\langle S_i^{(t)}S_j^{(t^\prime)}\rangle\biggr].\label{eq:app_Z_derivative_psi2_original}
\end{align}
Hence, by comparing \eqref{eq:app_Z_derivative_psi2} and \eqref{eq:app_Z_derivative_psi2_original} and taking the average over $i$ and $j$ with Kronecker's delta $\delta_{i,j}$, we obtain
\begin{align}
    q^{(t,t^\prime)}=\mathbb{E}_{\{\bm{\xi}^\mu\}}\left[\left\langle\frac{1}{N}\sum_{i=1}^NS_i^{(t)}S_i^{(t^\prime)}\right\rangle\right]
\end{align}

\paragraph{Response function $K^{(t,t^\prime)}$: }

Following eq.\eqref{eq:bar-Z-in-lemma}, we obtain
\begin{align}
	&\lim_{\bm{\psi}\to\bm{0}} 
	\frac{\partial \bar{Z}[\bm{\psi}]}
	{\partial \psi_i^{(t)}} 
	= 
	-\rmi \langle S^{(t)} \rangle_i, 
    \label{eq:app_Z_derivative_psi}\\
    &
	\lim_{\bm{\psi}\to\bm{0}} 
	\frac{\partial \bar{Z}[\bm{\psi}]}
	{\partial \theta_i^{(t)}} 
	= -\rmi\langle \hat{h}^{(t)}\rangle_i.
    \label{eq:app_Z_derivative_theta}\\
    	& 
	\lim_{\bm{\psi}\to\bm{0}} \!
	\frac{\partial^2 \bar{Z}[\bm{\psi}]}
	{\partial \psi_i^{(t)} \partial \theta_{j}^{(t')}}
	\!=\! 
	-\delta_{i,j} 
	\langle S^{(t)} \hat{h}^{(t')}\rangle_i 
	\!-\! (1\!-\!\delta_{i,j}) 
	\langle S^{(t)}\rangle_i 
	\langle \hat{h}^{(t')}\rangle_{j}.
	\label{eq:app_Z_derivative_psi_theta}
\end{align}
Meanwhile, following the definition of $\overline{Z}[\bm{\psi}]$ (eq.\eqref{eq:def-Z}), we obtain
\begin{align}
        &\lim_{\bm{\psi}\to\bm{0}}\frac{\partial\overline{Z}[\bm{\psi}]}{\partial\psi_i^{(t)}}=-i\mathbb{E}_{\{\bm{\xi}^\mu\}}\biggl[\langle S_i^{(t)}\rangle\biggr],\label{eq:app_Z_derivative_psi_original}\\
        &\lim_{\bm{\psi}\to\bm{0}} 
	\frac{\partial \bar{Z}[\bm{\psi}]}
	{\partial \theta_i^{(t)}} 
	= 0\quad \mbox{for all } i,t
    \label{eq:app_Z_derivative_theta_zero}\\
    &\lim_{\bm{\psi}\to\bm{0}} \!
	\frac{\partial^2 \bar{Z}[\bm{\psi}]}
	{\partial \psi_i^{(t)} \partial \theta_{j}^{(t')}}
	\!=\! -\rmi\mathbb{E}_{\{\bm{\xi}^\mu\}}\biggl[\frac{\partial}{\partial\theta_j^{(t^\prime)}}\langle S_i^{(t)}\rangle\biggr],
    \label{eq:app_Z_derivative_psi_theta_original}
\end{align}
where we use relationships $\overline{Z}[\bm{\psi}=\bm{0}]=1$ in deriving eq.\eqref{eq:app_Z_derivative_theta_zero}.
From the correspondence between eqs.\eqref{eq:app_Z_derivative_theta} and \eqref{eq:app_Z_derivative_theta_zero},
eq.\eqref{eq:app_Z_derivative_psi_theta} is simplified as
\begin{align}
    	\lim_{\bm{\psi}\to\bm{0}} \!
	\frac{\partial^2 \bar{Z}[\bm{\psi}]}
	{\partial \psi_i^{(t)} \partial \theta_{j}^{(t')}}
	\!=\! 
	-\delta_{i,j} 
	\langle S^{(t)} \hat{h}^{(t')}\rangle_i, 
    \label{eq:app_Z_derivative_psitheta_centered}
\end{align}
combining with \eqref{eq:app_Z_derivative_psi_theta}.
From the correspondence between eqs.\eqref{eq:app_Z_derivative_psi_theta_original} and \eqref{eq:app_Z_derivative_psitheta_centered},
we obtain
\begin{align}
    K^{(t,t^\prime)}=
\frac{\rmi}{N}\sum_{j=1}^N\mathbb{E}_{\{\bm{\xi}^\mu\}}\biggl[\frac{\partial}{\partial\theta_j^{(t^\prime)}}\langle S_j^{(t)}\rangle\biggr].
\end{align}

Here, R.H.S. represents the derivative of the state at time $t$ with respect to an external field applied at time $t^\prime$. 
Therefore, $K(t,t^\prime)$ can be nonzero only for $t>t^\prime$, considering causality.
\footnote{In the Modern Hopfield model, this causality property affects the saddle-point equation for $\{\hat{q}^{(t,t^\prime)}\}$. 
In the present model, however, it does not, because the saddle point is not characterized by $\{K^{(t,t^\prime)}\}$.}

\paragraph{Two-time correlation of response fields $Q$: }

Using \eqref{eq:bar-Z-in-lemma}, we obtain
\begin{align}
\lim_{\bm{\psi}\to\bm{0}}
\frac{\partial^2 \overline{Z}[\bm{\psi}]}
{\partial \theta_i^{(t)} \partial \theta_j^{(t')}}
=
-\delta_{i,j}
\langle \widehat{h}^{(t)}\widehat{h}^{(t')}\rangle_i.
\label{eq:app_Z_derivative_theta2}
\end{align}
On the other hand, from the definition of $\overline{Z}[\bm{\psi}]$, we have
\begin{align}
\lim_{\bm{\psi}\to\bm{0}}
\frac{\partial^2 \overline{Z}[\bm{\psi}]}
{\partial \theta_i^{(t)} \partial \theta_j^{(t')}}
=
0
\quad\text{for all }i,j,t,t',
\label{eq:app_Z_derivative_theta2_zero}
\end{align}
Comparing \eqref{eq:app_Z_derivative_theta2} with
\eqref{eq:app_Z_derivative_theta2_zero}, we conclude that
$Q^{(t,t')}=0$ for all $t$ and $t'$ at the saddle point.

\paragraph{Order parameter for retrieval $m$: }

We introduce a modified generating functional
\begin{align}
    \overline{Y}_{i\mu}[\bm{\psi}]=\mathbb{E}_{\bm{\xi}}\biggl[\xi_i^{\mu}\sum_{\bm{S}^{0:T}}p[\bm{S}^{0:T}]\exp(-\rmi\sum_{t=0}^{T-1}\bm{S}^{(t)}\cdot\bm{\psi}^{(t)})\biggr].
    \label{eq:app_def_Y}
\end{align}
Using this function, we obtain
\begin{align}
    &\lim_{\bm{\psi}\to\bm{0}}\frac{\partial\overline{Y}_{i\mu}[\bm{\psi}]}{\partial\psi_i^{(t)}}=-\rmi\mathbb{E}_{\{\bm{\xi}^\mu\}}\biggl[\xi_i^\mu\langle S_i^{(t)}\rangle\biggr].\label{eq:app_Y_derivative_psi_original}
\end{align}
Applying the same calculation to $\overline{Y}_{i\mu}[\bm{\psi}]$ as was used for $\overline{Z}[\bm{\psi}]$ in Proposition~\ref{prop:Z_form}, we obtain
\begin{align}
    \lim_{\bm{\psi}\to\bm{0}}\frac{\partial\overline{Y}_{i\mu}[\bm{\psi}]}{\partial\psi_i^{(t)}}=-\rmi\langle \xi_i^\mu S_i^{(t)}\rangle_i,
\end{align}
and therefore, at the saddle point,
\begin{align}
    m^{(t)}=\frac{1}{N}\sum_{i=1}^N\langle \xi_i^t S_i^{(t)}\rangle_i.
\end{align}

\paragraph{Auxiliary order parameter $k$: }

Using \eqref{eq:app_def_Y}, we obtain
\begin{align}
\lim_{\bm{\psi}\to\bm{0}}
\frac{\partial\overline{Y}_{i\mu}[\bm{\psi}]}
{\partial\theta_i^{(t)}}
=0.
\end{align}
On the other hand, we also obtain
\begin{align}
\lim_{\bm{\psi}\to\bm{0}}
\frac{\partial\overline{Y}_{i\mu}[\bm{\psi}]}
{\partial\theta_i^{(t)}}
=-\rmi\langle\xi_i^\mu\widehat{h}_i^{(t)}\rangle_i.
\end{align}
Therefore,
$\langle\xi_i^\mu\widehat{h}_i^{(t)}\rangle_i=0$ for all $i$ and $\mu$, and hence
\begin{align}
    k^{(t)}=0.
\end{align}

\subsection{Saddle points of $\widehat{\cal Q}$ and effective distribution}

The saddle point of $\widehat{\cal Q}$ is given by
\begin{align}
    \widehat{m}^{(t)}&=k^{(t)}n(n-1)(m^{(t)})^{n-2}\\
    \widehat{q}^{(t,t^\prime)}&=-\rmi\frac{n^2\alpha_n Q^{(t,t^\prime)}}{2}\sum_{k=0}^{n-1}A(n-1,k)k(q^{(t,t^\prime)})^{k-1}\\
    \widehat{k}^{(t)}&=n(m^{(t)})^{n-1}\\
    \widehat{Q}^{(t,t^\prime)}&=- \rmi \frac12 R^{(t,t')}\\
    \widehat{K}^{(t,t^\prime)}&=0.
\end{align}
Since $k^{(t)}=0$ and $Q^{(t,t^\prime)}=0$, it follows immediately that
$\widehat{m}^{(t)}=\widehat{q}^{(t,t^\prime)}=0$.

\subsection{Effective noise distribution}

Among ${\cal Q}$, non-zero quantities are $m^{(t)}$, $q^{(t,t^\prime)}$ and $K^{(t,t^\prime)}$.
However, substituting $\widehat{K}^{(t,t^\prime)}=0$ into \eqref{eq:bar-Z-in-lemma} removes all dependence on $K^{(t,t^\prime)}$. Hence, although $K^{(t,t^\prime)}$ is nonzero, it does not contribute to the dynamics.
Therefore, the dynamics is completely characterized by
$m^{(t)}$ and $q^{(t,t^\prime)}$.
As a consequence, the effective Hamiltonian
\eqref{eq:eff_H}, which describes temporal correlations,
reduces to
\begin{align}
        {\cal H}_{i}&=-\frac{\rmi}{2} 
		\hat{\bm{h}}_i^\top\underline{R}\hat{\bm{h}}_i +\underline{\bm{\psi}}_i^\top\underline{\bm{S}}_i-\sum_{t=0}^{T-1}\hat{h}_i^{(t)}
		( h_i^{(t)} - \mu_i^{(t)} ),
        \label{eq:app_eff_H}
\end{align}
where $\mu_i^{(t)}=n(m^{(t)})^{n-1}\xi_i^{t+1} + \theta_i^{(t)}$.

To evaluate $\langle f(S)\rangle_i$ with $\xi_i^{(t)}=\xi^{(t)}$ and $\theta_i^{(t)}=\theta^{(t)}$,
we integrate out the auxiliary variables
$\bm{h}_i$ and $\hat{\bm h}_i$,
which appear only through
$\omega_i$.
From \eqref{eq:def_omega}, we obtain
\begin{align}
\nonumber
    &\int d\bm{h}_id\hat{\bm{h}}_i\omega_i(\underline{\bm{S}}_i,\underline{\bm{h}}_i,\widehat{\underline{\bm{h}}}_i;\widehat{\cal Q},\underline{\bm{\psi}}_i)\\
    \nonumber
    &=p[S_i^{(0)}]\int d\bm{h}_i
			\prod_{t=0}^{T-1}  
			\delta [S_i^{(t+1)} ; \mathrm{sgn}(h_i^{(t)}) ]\exp(-i  \underline{\bm{\psi}}_i^\top\underline{\bm{S}}_i)\\
    &\times\frac{1}{\sqrt{(2\pi)^{T}|\underline{R}|}}\exp \biggl(
        -\frac{1}{2}(\underline{\bm{h}}_i-\underline{\bm{\mu}}_i)^\top\underline{R}^{-1}(\underline{\bm{h}}_i-\underline{\bm{\mu}}_i)\biggl).
        \label{eq:omega_integral}
\end{align}
Using this expression, we can show that ${\cal Z}_i(\widehat{{\cal Q}}^*,\underline{\bm{\psi}}_i=\bm{0})=1$.
Performing the remaining integration with respect to $\bm{h}_i$ in the numerator of Eq.~\eqref{eq:single_unit_measure}, we immediately obtain Eq.~\eqref{eq:def_average}.

\section{Large-$n$ asymptotics}
\label{sec:app_abs_c}
To obtain $m_c$ and $\alpha'_{c,n}$ satisfying Eqs.~\eqref{eq:crit_m} and \eqref{eq:crit_deriv}, we analyze these equations asymptotically for large $n$.
The derivative in Eq.~\eqref{eq:crit_deriv} can be evaluated exactly, which gives
\begin{align}
    1=\sqrt{\frac{2}{\pi}}\,\frac{(n-1)m_c^{\,n-2}}{\sqrt{\alpha'_{c,n}}}
    \exp\left[-\frac{m_c^{\,2(n-1)}}{2\alpha'_{c,n}}\right].
    \label{eq:large_n_tangency}
\end{align}
To treat Eq.~\eqref{eq:crit_m}, we use the asymptotic expansion of the error function for large $z$,
\begin{align}
    \operatorname{erf}(z)=1-\frac{e^{-z^2}}{\sqrt{\pi}\,z}\left[1+O(z^{-2})\right],
\end{align}
which is applicable here because the argument $z_c=m_c^{\,n-1}/\sqrt{2\alpha'_{c,n}}$ satisfies $z_c^2\simeq\ln n\to\infty$, as confirmed below.
Applying this expansion to Eq.~\eqref{eq:crit_m}, we obtain
\begin{align}
    1-m_c=
    \frac{\sqrt{2\alpha'_{c,n}}}{\sqrt{\pi}\,m_c^{\,n-1}}
    \exp\left[-\frac{m_c^{\,2(n-1)}}{2\alpha'_{c,n}}\right]
    \left[1+o(1)\right].
    \label{eq:large_n_fixed_point}
\end{align}
Eliminating the exponential factor using Eq.~\eqref{eq:large_n_tangency}, we find
\begin{align}
    1-m_c=\frac{\alpha'_{c,n}}{(n-1)m_c^{\,2n-3}}\left[1+o(1)\right],
    \label{eq:large_n_mc_relation}
\end{align}
which implies $1-m_c=O(\alpha'_{c,n}/n)$.
Consequently, $(n-2)\ln m_c=O(\alpha'_{c,n})$ and $m_c^{\,2(n-1)}=1-2\alpha'_{c,n}[1+o(1)]$.

To determine $\alpha'_{c,n}$, we take the logarithm of Eq.~\eqref{eq:large_n_tangency}:
\begin{align}
    \frac{m_c^{\,2(n-1)}}{2\alpha'_{c,n}}
    =\ln(n-1)+(n-2)\ln m_c-\frac{1}{2}\ln\alpha'_{c,n}+\frac{1}{2}\ln\frac{2}{\pi}.
    \label{eq:large_n_log}
\end{align}
By the estimates above, the left-hand side equals $1/(2\alpha'_{c,n})+O(1)$, and the second term on the right-hand side is $o(1)$.
To leading order, this gives $1/(2\alpha'_{c,n})\simeq\ln n$; substituting it into the term $-\frac{1}{2}\ln\alpha'_{c,n}$ yields
\begin{align}
    \frac{1}{2\alpha'_{c,n}}=\ln n+\frac{1}{2}\ln\ln n+O(1),
\end{align}
and hence
\begin{align}
    \alpha'_{c,n}=\frac{1}{2\ln n}\left[1+O\left(\frac{\ln\ln n}{\ln n}\right)\right].
    \label{eq:large_n_alpha}
\end{align}
The critical overlap then follows from Eq.~\eqref{eq:large_n_mc_relation}.
Since $1-m_c=O(\alpha'_{c,n}/n)$, the factor in its denominator satisfies $m_c^{\,2n-3}=1+O(\alpha'_{c,n})=1+O(1/\ln n)$, and $(n-1)^{-1}=n^{-1}[1+O(n^{-1})]$.
Substituting Eq.~\eqref{eq:large_n_alpha}, we obtain
\begin{align}
    m_c=1-\frac{1}{2n\ln n}\left[1+O\left(\frac{\ln\ln n}{\ln n}\right)\right].
    \label{eq:large_n_mc}
\end{align}

\nocite{*}

\bibliography{reference}

\end{document}